\documentclass[10pt]{article}

\usepackage[reqno,tbtags]{amsmath}
\usepackage{epsfig}
\usepackage{cite}% replaces [1,2,3] by [1-3]

\numberwithin{equation}{section}

\newcommand{\ep}{\epsilon}

\newcommand{\bq}{\begin{equation}}
\newcommand{\eq}{\end{equation}}

\newcommand{\ssL}{{\scriptscriptstyle{L}}}
\newcommand{\ssN}{{\scriptscriptstyle{N}}}
\newcommand{\ssO}{{\scriptscriptstyle{O}}}

\begin{document}
%----------------
\begin{titlepage} %suppresses the first page number count; so page 1 is after the title
%----------------
\begin{flushleft}
%{\small\tt DESY 06-154\\[-2mm]SFB/CPP-06-40}\\[3mm]
{\normalsize \tt
DESY 07-053 \\ 
SFB/CPP-07-15 \\
HEPTOOLS 07-010
\\
%\timestamp
%Draft, sent out Friday, 2 March 2007 
}
\end{flushleft}

\vspace{1cm}

\begin{center}
\large\bf
{\LARGE\bf 
Two-Loop Fermionic Corrections
\\ [.3cm] 
 to Massive Bhabha Scattering}\\[1cm]
%\date{\today}
\rm

{ Stefano Actis$^{a}$},~
{ Micha{\l} Czakon$^{b,c}$},~
{ Janusz Gluza$^{d}$},~
{ Tord Riemann$^{a}$}\\[.7cm]

%{ Stefano Actis$^{a}$}\\[.3cm] { Michal Czakon$^{b}$}\\[.3cm] { Janusz Gluza$^{c}$}\\[.3cm] { Tord Riemann$^{a}$}\\[.7cm]
{\em 
$^{a}$Deutsches Elektronen-Synchrotron, DESY,\\
Platanenallee 6, D-15738 Zeuthen, Germany \\[.5cm] 
$^{b}$Institut f\"ur Theoretische Physik und Astrophysik, Universit\"at W\"urzburg,\\
Am Hubland, D-97074 W\"urzburg, Germany \\[.5cm] 
$^{c}$Institute of Nuclear Physics, NCSR ``DEMOKRITOS'',\\
15310 Athens, Greece \\[.5cm]
$^{d}$Institute of Physics, University of Silesia,\\
Uniwersytecka 4, PL-40007 Katowice, Poland} 
\end{center}

\normalsize

\vspace{1cm}

\begin{abstract}
We evaluate the two-loop corrections to Bhabha scattering from fermion loops in the context of pure Quantum Electrodynamics.
The differential cross section is expressed by a small number of 
Master Integrals with exact dependence on the fermion masses $m_e,m_f$ and the Mandelstam invariants $s,t,u$.
We determine the limit of fixed scattering angle and high energy,
assuming the hierarchy of scales $m_e^2 \ll m_f^2 \ll s,t,u$.
The numerical result is combined with the available non-fermionic contributions.
As a by-product, we provide an independent check of the known electron-loop contributions.
\end{abstract}
%\maketitle: gives format to \title \author \abstract
%--------------
\end{titlepage}
%--------------

%-----------------------
\section{Introduction}
%-----------------------
%
% --> This is part of the old introduction.
%
%Bhabha scattering is employed to measure the luminosity at electron-positron colliders 
%because of its clear experimental signature.
%Small-angle Bhabha scattering provides an efficient tool at high-energy ($\sim$ $100$ %$\rm{GeV}$) colliders.
%The large-angle kinematic region is relevant at machines operating at a $1$-$10$ %$\rm{GeV}$ center-of-mass energy.
%Furthermore, the luminosity spectrum at a future International Linear Collider
%will not be monochromatic due to beam-beam effects,
%and large-angle Bhabha scattering could disentangle the luminosity spectrum itself~\cite{Toomi:1997ke,Aguilar-Saavedra:2001rg}.
%The luminosity is determined through the ratio of the number of observed events
%and the theoretical prediction for the cross section.
%Therefore, a precise computation of radiative corrections to Bhabha scattering can minimize the luminosity error.
%
%The electroweak one-loop effects were computed a long time ago in~\cite{Consoli:1979xw}.
%In recent years, several authors have been focusing on two-loop contributions in pure Quantum Electrodynamics.
%
Bhabha scattering is one of the processes at $e^+e^-$ colliders with the highest experimental precision 
and represents an important monitoring process.
A notable example is its expected role for the luminosity determination at the future International Linear Collider ILC
by measuring small-angle Bhabha-scattering events at center-of-mass energies ranging from about 100 GeV (Giga-Z collider option) to several TeV.
Moreover, the large-angle region is relevant at colliders operating at 1--10 GeV.
For some applications a full two-loop calculation of the QED contributions is mandatory\footnote{
Note that leading two-loop effects in the electroweak Standard Model were already
incorporated in~\cite{Bardin:1991xe}.}.

A large class of QED two-loop corrections was determined in the seminal work of~\cite{Arbuzov:1998du}.
Later, the complete two-loop corrections in the limit of zero electron mass were obtained in~\cite{Bern:2000ie} thanks to the fundamental results of~\cite{Smirnov:1999gc,Tausk:1999vh}.
However, this result cannot be immediately applied, since the available Monte-Carlo 
programs (see e.g.~\cite{Berends:1983fs,Berends:1987jm,Jadach:1996is,Jadach:1995nk,Arbuzov:1997pj,Arbuzov:2005pt,CarloniCalame:2000pz,Balossini:2006wc}) employ a small, but non-vanishing electron mass.
The $\alpha^2 \ln(s/m_e^2)$ terms due to double boxes were derived from~\cite{Bern:2000ie} by 
the authors of~\cite{Glover:2001ev},
and the close-to-complete two-loop result in the ultra-relativistic limit was finally obtained 
in~\cite{Penin:2005kf,Penin:2005eh}. 
Note that the diagrams with fermion loops have  not been covered by this approach.
The virtual and real components involving electron loops could be added 
exactly in~\cite{Bonciani:2004gi,Bonciani:2004qt}.
The non-approximated analytical expressions for all two-loop corrections, 
except for double-box diagrams and for those with loops from heavier-fermion generations, 
can be found in~\cite{Bonciani:2005im}.
For a comprehensive investigation of the full set of the massive two-loop QED corrections,
including double-box diagrams, we refer to~\cite{Czakon:2004wm,Czakon:2006hb,Czakon:2006pa}.
The evaluation of the contributions from massive non-planar double box diagrams remains open so far.

In order to add another piece to the complete two-loop prediction for the Bhabha-scattering cross section in QED,
we evaluate here the so-far lacking diagrams containing  heavy-fermion loops.
The cross section correction is expressed by a small number of scalar Master Integrals,
where the \emph{exact} dependence on the masses of the fermions and the Mandelstam
variables $s$, $t$ and $u$ is retained.
In a next step, we assume a hierarchy of scales, $m_e^2 \ll m_f^2 \ll s,t,u$, 
where $m_e$ is the electron mass and $m_f$ is the mass of a heavier fermion. 
We derive explicit results neglecting terms suppressed by positive powers
of $m_e^2\slash m_f^2$, $m_e^2\slash x$ and $m_f^2\slash x$, where $x = s,t,u$.
This high-energy approximation describes the influence of muons and $\tau$ leptons and 
proves well-suited for practical applications. 
In addition, we provide an independent cross-check of the 
exact analytical results of~\cite{Bonciani:2004gi} (we
  used the files provided at \cite{Bonciani:web} for comparison) for $m_f =
  m_e$. 

The article is organized as follows.
In Section~\ref{exp} we introduce our notations and outline the  calculation and
in Section~\ref{main} we discuss the solution for each class of diagrams.
In Section~\ref{tot} we reproduce the complete result for the corrections from heavier fermions 
in analytic form and perform the numerical analysis. 
Section~\ref{final} contains the summary, and additional material on the Master Integrals is collected in the Appendix.

%-----------------------------------------
\section{Expansion of the Cross Section}\label{exp}
%-----------------------------------------
%--
We consider the Bhabha-scattering process,
%--
\bq
e^{_-} \, (p_1)\, +\, e^{_+} \, (p_2)\,
\to\,
e^{_-} \, (p_3)\, +\, e^{_+} \, (p_4),
\eq
%--
and introduce the Mandelstam invariants $s$, $t$ and $u$,
%--
\begin{eqnarray}\label{Mandelstam}
s &= &  \left(\, p_1+p_2\, \right)^2 = 4 \, E^2, 
%\nonumber 
\\
t &= & \left(\, p_1-p_3\, \right)^2 =  -4\left(\,E^2\,-\,m_e^2\,\right) \,\sin^2 \frac{\theta}{2}, %\nonumber 
\\
u &= & \left(\, p_1-p_4\, \right)^2 =  -4\left(\,E^2\,-\,m_e^2\,\right) \,\cos^2 \frac{\theta}{2},
\end{eqnarray}
%--
where $m_e$ is the electron mass,
$E$ is the incoming-particle energy in the center-of-mass frame
and $\theta$ is the scattering angle.
In addition,
$
 s\, +  \, t \, +\, u\, = \, 4\, m_e^2.$
%--
%--

In the kinematical region $m_e^2 \ll s,t,u$ 
the leading-order (LO) differential cross section with
respect to  the solid angle $\Omega$ reads as
%--
\begin{equation}\label{born}
\frac{d \sigma^{\rm{\ssL\ssO}}}{d \Omega}\,
=\,
\frac{\alpha^2}{s}\,
\Bigl[\,
\frac{1}{s^2}\,
\Bigl(\, \frac{s^2}{2}\, + \,  t^2\, +\, s\,t \,\Bigr)\,
+\,
\frac{1}{t^2}\,
\Bigl(\, \frac{t^2}{2}\, + \,  s^2\, +\, s\,t \,\Bigr)\,
+\,  \frac{1}{s\, t}\, \left(\, s\, +\, t\, \right)^2\,
\Bigr],
\end{equation}
%--

\noindent
where $\alpha$ is the fine-structure constant.
% and we neglected terms suppressed by positive powers of $m_e^2\slash s$ and $m_e^2\slash t$.
At higher orders in perturbation theory we write an expansion
in $\alpha$,
%--
\bq\label{expansion}
\frac{d \sigma}{d \Omega}\,=\,
\frac{d \sigma^{\rm{\ssL\ssO}}}{d \Omega}\,+\,
\left(\frac{\alpha}{\pi}\right)
\frac{d \sigma^{\rm{\ssN\ssL\ssO}}}{d \Omega}\,+\,
\left(\frac{\alpha}{\pi}\right)^2
\frac{d \sigma^{\rm{\ssN\ssN\ssL\ssO}}}{d \Omega}
+ {\cal O}(\alpha^5).
\eq
%--
Here $d \sigma^{\rm{\ssN\ssL\ssO}}$ and
$d \sigma^{\rm{\ssN\ssN\ssL\ssO}}$
summarize the next-to-leading order (NLO) 
and next-to-next-to-leading order (NNLO)
corrections to the differential cross section.
In the following it will be understood that we consider only components
generated by diagrams containing one or two fermion loops.

%%%%%%%%%%%%%%
\begin{figure}[t]
\begin{center}
\includegraphics[scale=0.6]{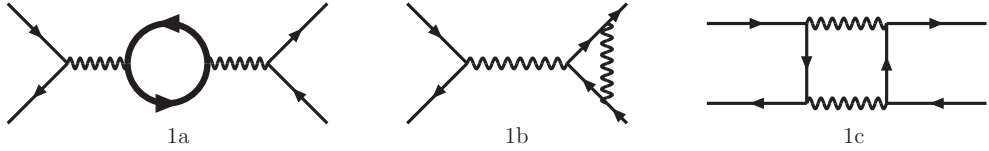}
\end{center}
\caption{
Classes of Bhabha-scattering one-loop diagrams.
A thin fermion line represents an electron, a thick one
can be any fermion.
The full set of graphs can be obtained through
proper permutations.
We refer to~\cite{webPage:2006xx} for the reproduction of the full set of graphs.
%We refer to~\cite{Bonciani:2004gi} for a detailed discussion.
}
\label{1loop}
\end{figure}
%%%%%%%%%%%%%%

\subsection{NLO Differential Cross Section}
\label{conventions}
%--
The NLO term follows from the 
interference of the one-loop vacuum-polarization diagrams of class \rm{1a}
(see Figure~\ref{1loop})
with the tree-level amplitude,
\begin{eqnarray}\label{NLO}
\frac{d \sigma^{\rm{\ssN\ssL\ssO}}}{d \Omega}\,=\,
\frac{d \sigma^{\rm{1a}\times\rm{tree}}}{d \Omega}=
\frac{\alpha^2}{s}
&\Bigl\{&
\frac{1}{s^2}\,
\Bigl(\, \frac{s^2}{2}\, + \,  t^2\, +\, s\,t \,\Bigr)\,
\, 2\, \sum_{f}\, Q_f^2\, \text{Re}\, \Bigl[\,   \Pi^{(1)}_f(s)\, \Bigr] \nonumber \\
&+& \frac{1}{t^2}\, 
\Bigl(\, \frac{t^2}{2}\, +\, s^2\, +\, s\,t \,\Bigr)\,
\, 2\,\sum_{f}\, Q_f^2\ \text{Re}\, \Bigl[ \,   \Pi^{(1)}_f(t)\, \Bigr] \nonumber \\
&+& \frac{1}{s\, t}\, \left(\, s\, +\, t\, \right)^2\,
\,  \sum_{f}\, Q_f^2\, \text{Re}\, \Bigl[\,
 \Pi^{(1)}_f(s)\,+\, \Pi^{(1)}_f(t)\, 
\Bigr]\,\,\,
\Bigr\}.
\end{eqnarray}
%--
Here $\Pi^{(1)}_f(x)$ is the  renormalized one-loop vacuum-polarization function and
the sum over $f$ runs over the massive fermions, e.g. the electron ($f=e$), the muon ($f=\mu$), the $\tau$ lepton ($f=\tau$).
%and the top quark ($f=t$).
$Q_f$ is the electric-charge quantum number, $Q_f=-1$ for leptons.
% and$Q_f=2\slash 3$ for the top quark.

In this paper we will focus on asymptotic expansions in the high-energy limit.
In order to fix our normalizations explicitly, we reproduce here the
exact result for $\Pi^{(1)}_f(x)$ in dimensional regularization.
Adding $\Pi^{(1)ct}_f(x)$, the counterterm contribution in the on-mass-shell scheme (see the following
discussion in Subsection~\ref{reno}),
to $\Pi^{(1)un}_f(x)$, the unrenormalized one-loop vacuum polarization function, we get
%--
\begin{eqnarray}\label{Pi10ex}
\Pi^{(1)}_f(x)
&=&
\Pi^{(1)un}_f(x) + \Pi^{(1)ct}_f(x), %\nonumber
\\\label{Pi10exa}
\Pi^{(1)un}_f(x) &=&
\frac{1}{2(D-1)} \left[ 2(D-2)\frac{1}{x}A_0(m_f)  - \left(D-2+4\frac{m_f^2}{x}\right) B_0(x,m_f)%\texttt{SE2l2m} 
 \right]  , %\nonumber
\\\label{Pi10exb}
\Pi^{(1)ct}_f(x) &=& \frac{1}{3}F_\ep \left( \frac{m_e^2}{m_f^2}\right)^\ep \left( \frac{1}{\ep} + \frac{\zeta_2}{2}\ep\right),
\end{eqnarray}
%--
where $\epsilon =(4-D)\slash 2$ and $D$ is the number of space-time dimensions.
The normalization factor is 
\bq
F_\epsilon\, = \,
\, \left(\, \frac{m_e^2\, \pi\, e^{\gamma_E}}{\mu^2 }\, \right)^{-\epsilon},
\eq
%--
$\mu$ is the 't Hooft mass unit and $\gamma_E$ is the
Euler-Mascheroni constant.
Standard one-loop integrals appearing in Eq.~\eqref{Pi10ex} are defined by
\begin{eqnarray}
A_0(m)&=&\frac{\mu^{4-D}}{i \pi^2} \int d^Dk \frac{1}{k^2-m^2},%\nonumber
\\
B_0(p^2,m)&=&\frac{\mu^{4-D}}{i \pi^2} \int d^Dk \frac{1}{(k^2-m^2)[(k+p)^2-m^2]}.
\end{eqnarray}
Note that Master Integrals with \texttt{l} lines and an internal scale $m$ were derived in~\cite{Czakon:2004wm,webPage:2006xx}
setting $m=1$.  For the present computation we introduce a scaling by a factor  $m_f^{D-2l}$
and we get
 \begin{eqnarray}\label{Pi10exc}
A_0(m_f) &=& F_\ep\, \left(\frac{m_e^2}{m_f^2}\right)^\ep ~\, m_f^2 ~\, \texttt{T1l1m}, %\nonumber
\\
B_0(x,m_f) &=& F_\ep\, \left(\frac{m_e^2}{m_f^2}\right)^\ep  \texttt{SE2l2m[x]}. 
\end{eqnarray}
In the small-mass limit, $A_0$ vanishes (the result for \texttt{T1l1m} can be read in Eq.(4) of~\cite{Czakon:2004wm}),
and the one-loop self-energy\footnote{Here, the argument $x$ of  \texttt{SE2l2m[x]} is one of the relativistic invariants $s,t,u$. This deviates from earlier conventions, where we denoted by $x$  the dimensionless conformal transform of $s,t,u$. This remark applies also to Master Integrals in the Appendix.} reads as
\begin{equation}\label{selfOneOld}
\texttt{SE2l2m[x]} = \frac{1}{\ep} + 2 +L_f(x) + \ep \left[
4 - \frac{\zeta_2}{2} + 2\, L_f(x) + \frac{1}{2}\, L^2_f(x) \right] \, ,
\end{equation}
where we introduced the short-hand notation for logarithmic functions
(in our conventions the logarithm has a cut along the negative real  axis),
%--
\bq\label{LOG1}
L_f(x)\, =\, \ln\left(\,-\,\frac{m_f^2}{x+\,i\,\delta}\,\right),\qquad \delta\to 0_+.
\eq
%--
Finally, neglecting ${\cal O}(m_f^2/x)$ terms, $\Pi^{(1)}_f(x)$ reads as
%--
%% corr. to exact result of eqs. 82,83 of Bonciani:2004gi,hep-ph/0405275
\bq\label{Pi10}
\Pi^{(1)}_f(x)\,=\,
- \, \frac{F_\ep}{3}\, \left(\frac{m_e^2}{m_f^2}\right)^\ep \, \left\{\,
\frac{5}{3}\,
+\, L_f(x)\,
+ \, \epsilon\, \left[\,
 \frac{28}{9}\,
- \, \zeta_2\,
+\, \frac{5}{3}\, L_f(x)\,
+\, \frac{1}{2}\, L_f^2(x)\, 
 \,\right]\,
\right\}
.
\eq
%--
Note that the ${\cal O}(\epsilon)$ term in Eq.~\eqref{Pi10} is not required for the NLO computation, 
but it will become relevant at NNLO.
Here $\Pi^{(1)}_f(x)$ will be
combined with infrared-divergent graphs
showing single poles in the $\epsilon$ plane for $\epsilon=0$.
The exact result for $\Pi^{(1)}_f(x)$ is
available at~\cite{webPage:2006xx}.

%%%%%%%%%%%%%%
\begin{figure}[t]
\begin{center}
\includegraphics[scale=0.6]{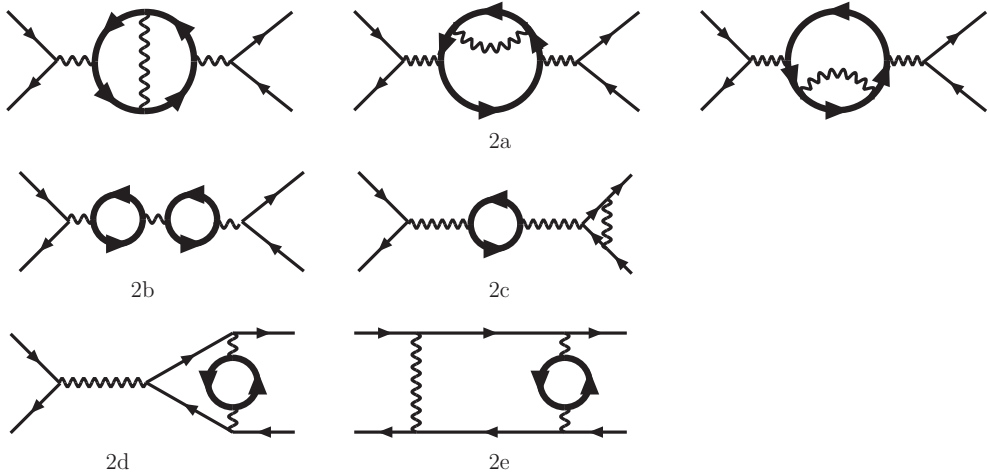}
\end{center}
\caption{
Classes of Bhabha-scattering two-loop diagrams containing at least one fermion loop.
We use the conventions of Figure~\ref{1loop}.
Note that class \rm{2a} contains three 
topologically different subclasses.
We refer to~\cite{webPage:2006xx} for the reproduction of the full set of graphs.
%We refer again to~\cite{Bonciani:2004gi} for a detailed discussion of the permutations required to obtain the full set of graphs.}
}\label{2loop}
\end{figure}
%%%%%%%%%%%%%%

\subsection{Outline of the NNLO Computation}
\label{outline}
%--
At NNLO we have to consider:
\begin{itemize}
\item
The interference of the two-loop
diagrams of classes \rm{2a}-\rm{2e} (see Figure~\ref{2loop})
with the tree-level amplitude;
\item
The interference of the one-loop vacuum-polarization diagrams of class \rm{1a}
with the full set of graphs of classes \rm{1a}-\rm{1c} (see Figure~\ref{1loop}).
\end{itemize}
The complete result can be organized as
\bq\label{complete}
\frac{d \sigma^{\rm{\ssN\ssN\ssL\ssO}}}{d \Omega}\,=\,
\underbrace{\sum_{\rm{i}=\rm{a},\ldots,\rm{e}} \frac{d \sigma^{\rm{2i}\times\rm{tree}}}{d \Omega}}_{\rm{2-loop}\times\rm{tree}}\,+\,
\underbrace{\sum_{\rm{i}=\rm{a},\ldots,\rm{c}}\frac{d \sigma^{\rm{1a}\times\rm{1i}}}{d \Omega}}_{\rm{1-loop}\times\rm{1-loop}}.
\eq

%%%%%%%%%%%%%%
\begin{figure}[t]
\begin{center}
\includegraphics[scale=0.6]{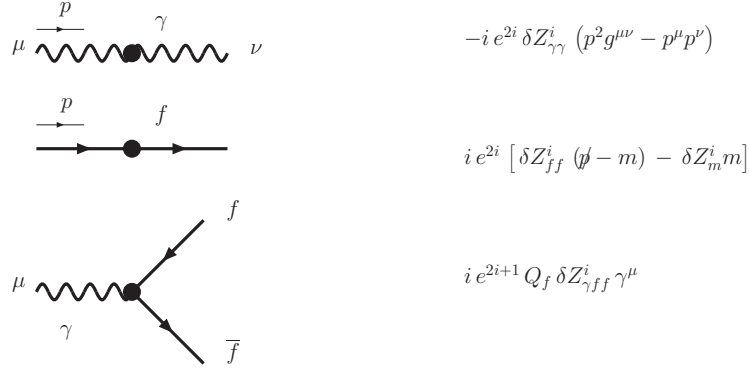}
\end{center}
\caption{
Counterterm-dependent Feynman rules relevant for Bhabha scattering
for $i=1$ (one loop) and $i=2$ (two loops).
Note that in the on-mass-shell scheme $e^2=4\pi\alpha$ at all orders
in perturbation theory.
}
\label{cts}
\end{figure}
%%%%%%%%%%%%%%
\noindent
In order to compute the NNLO differential cross section
we use the following {\em reduction strategy}:
%--
\begin{itemize}
%--
\item
The generation of all the diagrams is simple and has been made with the computer-algebra systems 
 \texttt{GraphShot}~\cite{Actis:gshot} and \texttt{qgraf/DIANA}~\cite{Nogueira:1993ex,Nogueira:1997??,Tentyukov:1999is}.
We spin-sum the squared matrix elements
and take the traces over Dirac indices in $D$ dimensions using the computer-algebra system \texttt{FORM}~\cite{Vermaseren:2000nd}.
The resulting expressions are combinations of algebraic coefficients depending on $s,t,u,m_e,m_f$ and $\ep$ 
and two-loop integrals with scalar products containing the loop momenta in the numerators.
An example showing the complexity of the result (two-loop box diagram of class {\rm 2e}, see Figure~\ref{2loop}) 
can be found at~\cite{webPage:2006xx}.
%--
\item
We reduce the loop integrals to a set of Master Integrals
by means of the \texttt{IdSolver} implementation~\cite{Czakon:2004uu2}
of the Laporta algorithm~\cite{Laporta:1996mq,Laporta:2001dd}.
The complete {\em list of massive Bhabha-scattering Master Integrals} can be found in~\cite{Czakon:2004wm}. 
%--
\end{itemize}
%--

Next, we {\em evaluate the Master Integrals}:
\begin{itemize}
\item
Integrals arising from graphs of classes \rm{1a}-\rm{1c} (Figure~\ref{1loop}),
 \rm{2a}-\rm{2c} (Figure~\ref{2loop}) and \rm{2d}-\rm{2e} (Figure~\ref{2loop}, with
electron loops) have been computed exactly through the method
of differential equations in the external kinematic variables
and expressed through Harmonic Polylogarithms~\cite{Remiddi:1999ew} or
Generalized Harmonic Polylogarithms~\cite{Gehrmann:1999as,Gehrmann:2000zt}.
Here we agree perfectly with the work of~\cite{Bonciani:2004gi,Bonciani:web}.
Non-approximated results for the various components of the differential cross section
are collected in a ~\texttt{Mathematica}~\cite{Wolfram} file at~\cite{webPage:2006xx}.
\item
Integrals generated by the diagrams of classes \rm{2d}-\rm{2e} (Figure~\ref{2loop},
with heavy-fermion loops) are computed through a method based on 
asymptotic expansions of Mellin-Barnes representations.
We derived appropriate Mellin-Barnes representations~\cite{Usyukina:1975yg,Boos:1991rg} for each Master Integral
and performed an analytic continuation in $\epsilon$ from a range where the integral is regular to 
the origin of the $\epsilon$ plane~\cite{Smirnov:1999gc,Tausk:1999vh}.
This is done by an automatic procedure implemented in the package \texttt{MB.m}~\cite{Czakon:2005rk}.
To proceed further, we assume a hierarchy of scales, $m_e^2 \ll m_f^2 \ll s,t,u$, where $f\neq e$.
After identifying the leading contributions in the fermion masses
(in the same spirit as in~\cite{Roth:1996pd}), 
we express the integrals by series over residua,
and the latter are sumed up analytically in terms of polylogs by means of the package \texttt{XSUMMER}\cite{Moch:2005uc}.
Asymptotic expansions for the master integrals with two different masses were given in~\cite{Actis:2006dj}.
They,  
and also few lacking expansions of simpler masters needed here have been collected in Appendix~\ref{appendix}.
We refer for a detailed discussion to~\cite{Czakon:2006pa}, where 
the technique was employed to derive approximated results for 
the massive Bhabha-scattering planar box master integrals.
All the mass-expanded masters may also be found in a~\texttt{Mathematica} file at~\cite{webPage:2006xx}.
\end{itemize}

\subsection{Renormalization}\label{reno}
%--
In the following we will always deal with ultraviolet-renormalized quantities.
After regularizing the theory using dimensional regularization~\cite{'tHooft:1972fi,Bollini:1972ui},
we perform renormalization in the on-mass-shell scheme. Here we relate 
all free parameters to physical observables:
\begin{itemize}
\item[--]
The electric charge coincides
with the value of the electromagnetic coupling,
as measured in Thomson scattering,
at all orders in perturbation theory; 
\item[--]
The squared fermion masses are identified 
with the real parts of the poles of the Dyson-resummed propagators;
\item[--]
Finally, field-renormalization constants are chosen
in order to cancel external wave-function corrections.
\end{itemize}
Counterterm-dependent Feynman rules are shown in Figure~\ref{cts}.
Note that the presence of infrared divergencies at NNLO requires to compute
one-loop counterterms including ${\cal O}(\epsilon)$ terms.
%
%In showing the results we introduce an overall factor
%\bq
%F_\epsilon\, \equiv\,
%\, \left(\, \frac{m_e^2\, \pi\, e^{\gamma_E}}{\mu^2 }\, \right)^{-\epsilon},
%\eq
%--
%where $\mu$ is the arbitrary 't Hooft mass unit and $\gamma_E$ is the
%Euler-Mascheroni constant.
%The counterterms will depend on logarithms of mass ratios,
%--
%\bq\label{LOG2}
%L(R_f)\, =\, \ln\, \left(\frac{m_e^2}{m_f^2}\right).
%\eq
%--
%--
\subsubsection*{One-Loop Counterterms}
%--
The one-loop counterterms read as
%--
\begin{eqnarray}
\delta Z^1_{\gamma\gamma}&=&
-\,\frac{F_\epsilon}{12 \,\pi^2}\,
\sum_f \, Q_f^2\, \Bigl( \frac{m_e^2}{m_f^2}\Bigr)^{\ep}
\Bigl(\,
\frac{1}{\epsilon}\,
+\, \frac{\epsilon}{2}\, \zeta_2\,
\Bigr),
\label{ct11}
\\
%--
\delta Z^1_{ff} &=& \delta Z^1_{m}= -\, \frac{F_\epsilon}{16 \,\pi^2}\, Q_f^2\,\Bigl( \frac{m_e^2}{m_f^2}\Bigr)^{\ep}\,
\left[\, \frac{3}{\epsilon}\,
+ \, 4\,
+\, \epsilon \, \left(\,
8\, +\, \frac{3}{2} \, \zeta_2\,
\right)\,
\right],
\label{ct12}
\\
%--
\delta Z^1_{\gamma ff} &=&  \delta Z^1_{ff} , 
\label{ct13}
\end{eqnarray}
%--
where the last equation follows from the U(1) QED Ward identity.
In the ultrarelativistic limit, the one-loop fermion-mass counterterm
is not needed, since it is always multiplied by the fermion mass.
Note however that the same counterterm  is relevant for the \emph{exact} computation.
%--
\subsubsection*{Two-Loop Counterterms}
%--
At the two-loop level we get
%--
\begin{eqnarray}
\delta Z_{\gamma\gamma}^2&=&\,
-\frac{F_\epsilon^2}{128 \, \pi^4}\, 
\sum_f\, Q_f^4\, \Bigl( \frac{m_e^2}{m_f^2}\Bigr)^{2\ep}
\Bigl(\,
\frac{1}{\epsilon}\,
+\, \frac{15}{2}\,
\Bigr),
\\
%--
%\begin{split}
\delta Z^2_{\gamma ee}\, &=& \, 
\frac{F_\epsilon^2}{128 \, \pi^4} \,
\Bigl[
\,
\frac{1}{2\, \epsilon}
\, +\,
\frac{947}{36}
\,-\,
16\, \zeta_2\,
 +\,
\sum_{f\neq e} Q_f^2 \Bigl( \frac{m_e^2}{m_f^2}\Bigr)^{2\ep}
\Bigl(
\,
\frac{1}{2\, \epsilon}\,
-\,
\frac{5}{12}\,
\,\Bigr)
\Bigr]. 
%\end{split}
\end{eqnarray}
%--
The result for $\delta Z^2_{\gamma ee}$ is obtained including
just fermion-loop diagrams and neglecting ${\cal O}(m_e^2\slash m_f^2)$ terms
for $f\neq e$.
The expression for $\delta Z_{\gamma\gamma}^2$
(as well as the one-loop counterterms of Eqs.~\eqref{ct11}-\eqref{ct13}), instead,
is exact, since it follows from the single-scale diagrams
of classes {\rm 2a-2b} of Figure~\ref{2loop}.
Finally, we observe that the two-loop counterterm with two fermion lines is not required,
since the use of an on-mass-shell renormalization removes external
wave-function factors.
%--
\section{Two-Loop Corrections}\label{main}
%--
In this Section we show our \emph{approximated} results for all the components of 
the NNLO differential cross section of Eq.~\eqref{expansion}.
%--
Our short-hand notation for logarithmic functions can be found in Eq.~\eqref{LOG1}.
In addition, we define two combinations of the Mandelstam invariants:
% in order to show our results in a compact form,
%--
\begin{eqnarray}\label{kinCOEFF}
v_1(x,y;\epsilon)\,&=&\,
x^2\,+\, 2\, y^2\, +\,2\,x\, y\,-\epsilon\,x^2,
%\nonumber 
\\
v_2(x,y;\epsilon)\,&=&\, 
(\, x\,+\,y\,)^2\,-\,\epsilon\,(\,x^2\,+\,y^2\,+x\,y\,),
\end{eqnarray}
%--
where $x (y)=s,t,u$.
Note that for $\ep = 0$ these functions are proportional to the kinematical factors 
appearing in the Born cross section of Eq.~\eqref{born} and
the NLO corrections of Eq.~\eqref{NLO}.
Moreover, we introduce a compact notation which will prove useful
in discussing box corrections in Subsection~\ref{BOXES} and
the \emph{complete} NNLO differential cross section in Section~\ref{tot},
%--
\begin{equation}\label{LOG2}
L(R_f)\, =\, \ln\, \left(\frac{m_e^2}{m_f^2}\right).
\end{equation}
%--
\subsection{Vacuum-Polarization Corrections}
%--
The interference of the vacuum-polarization diagrams of classes \rm{2a} and \rm{2b} 
with the tree-level amplitude can be written as
%--
\begin{eqnarray}\label{self}
\frac{d \sigma^{\rm{2i}\times\rm{tree}}}{d \Omega}\,=\,
\frac{\alpha^2}{s}
&\Bigl\{&
\frac{1}{s^2}\,\,
v_1(s,t;0)\,\,
A^{\rm{2i}}(s)\,
\, + \, 
 \frac{1}{t^2}\,\, v_1(t,s;0)\,
\,  A^{\rm{2i}}(t) \nonumber \\
 &+&
 \frac{1}{s\,t}\,\, v_2(s,t;0)  \,\,
\Bigl[\, A^{\rm{2i}}(s)\, +\, A^{\rm{2i}}(t)\, \Bigr]
\,\,\, \Bigr\},\qquad \rm{i}=\rm{a},\rm{b}.
\end{eqnarray}
%--
Here we introduced the auxiliary functions $A^{\rm{2a}}(x)$ and $A^{\rm{2b}}(x)$,
which are expressed through the renormalized one- and two-loop vacuum-polarization
functions $\Pi^{(1)}_f(x)$ (see Eq.~\eqref{Pi10} ) and $\Pi^{(2)}_f(x)$,
%--
\begin{eqnarray}
A^{\rm{2a}}(x)&=&
\sum_{f}\,
Q_f^4\,
\text{Re}\,\Bigl[ \, 
\Pi^{(2)}_f(x) \, 
\Bigr]
, \label{Q4}\\
%--
A^{\rm{2b}}(x)&=&
\sum_{f_1,f_2}\,
Q_{f_1}^2\, Q_{f_2}^2\,
\text{Re}\,\Bigl[ \, 
\Pi^{(1)}_{f_1}(x) \, \, \Pi^{(1)}_{f_2}(x)\, 
\Bigr]
\label{otherLoopByLoop}
,
\end{eqnarray}
%--
where the result for $\Pi^{(2)}_f(x)$ in the small fermion-mass limit reads as
%--
\bq\label{Pi2}
\Pi^{(2)}_f(x)\,=\,
-\, \frac{5}{24}\,
+\, \zeta_3
-\, \frac{1}{4}\, L_f(x)\,
.
\eq
%--
Note that ${\cal O}(\epsilon)$ terms in Eq.~\eqref{self} coming from the kinematical coefficients 
of Eq.~\eqref{kinCOEFF} can be safely neglected,
since both $\Pi^{(1)}_f(x)$ and $\Pi^{(2)}_f(x)$ are infrared-finite quantities.
%--
\subsection{Vertex Corrections}
%--
The contribution of reducible (irreducible) vertex corrections to the NNLO
differential cross section can be readily derived from diagrams
of classes \rm{2c} (\rm{2d}) in Figure~\ref{2loop},
\begin{eqnarray}\label{VERTEX2}
\frac{d \sigma^{\rm{2i}\times\rm{tree}}}{d \Omega}\,=\,2\,
\frac{\alpha^2}{s}
&\Bigl\{
&\frac{1}{s^2}\,
\Bigl[\,
v_1(s,t;\epsilon)\,
A_{\rm{V}}^{\rm{2i}}(s)\,
\, +\,
s^2\, 
A_{\rm{M}}^{\rm{2i}}(s)
\Bigr]
+
\frac{1}{t^2}\,
\Bigl[\,
v_1(t,s;\epsilon)\,
A_{\rm{V}}^{\rm{2i}}(t)\,
\, +\,
t^2\, A_{\rm{M}}^{\rm{2i}}(t)\,
\Bigr] \nonumber \\
&+&  \frac{1}{s\, t}\, 
\Bigl[\,
v_2(s,t;\epsilon)\,
 \Bigl(\, A_{\rm{V}}^{\rm{2i}}(s)\,+\,
A_{\rm{V}}^{\rm{2i}}(t)\,\Bigr)
\,+\,
\frac{3}{2}\,
\Bigl(\,
s^2\, A_{\rm{M}}^{\rm{2i}}(s)\,
+\,
t^2 \,A_{\rm{M}}^{\rm{2i}}(t)\,
\Bigr)\,\nonumber \\
&+&
2\, s\,t\,
 \Bigl(\, A_{\rm{M}}^{\rm{2i}}(s)\,+\,
A_{\rm{M}}^{\rm{2i}}(t)\,\Bigr)\,
\Bigr]\,
\Bigr\},\qquad \rm{i}=\rm{c},\rm{d}.
\end{eqnarray}
%--
\subsubsection*{Reducible diagrams} 
%--
The auxiliary functions $A_{\rm{V}}^{\rm{2c}}(x)$ and
$A_{\rm{M}}^{\rm{2c}}(x)$ are given by the product
of the renormalized one-loop vacuum-polarization function $\Pi^{(1)}_f(x)$ 
(expanded  in Eq.~\eqref{Pi10}
including ${\cal O}(\epsilon)$ terms)
and the renormalized one-loop vector and magnetic vertex form factors $F^{(1)}_{\rm{V}}(x)$
and $F^{(1)}_{\rm{M}}(x)$,
\bq
A_{\rm{I}}^{\rm{2c}}(x)\, = \,
\sum_{f}\,
Q_f^2\,
\text{Re}\, \Bigl[\, F^{(1)}_{\rm{I}}(x) \,  \Pi^{(1)}_f(x)\,  \Bigr],\qquad
\rm{I}\,=\rm{V,M}.
\eq
%--
The asymptotic expansion of $F^{(1)}_{\rm{V}}(x)$ is given by
%--
\bq\label{Vec1}
\begin{split}
F^{1}_{\rm{V}}(x)\,&=\,
-\, \frac{F_\ep}{2\epsilon}\,
\, \Bigl[
1\, +\, L_e(x)
\Bigr]
-\, 1\, +\, \frac{1}{2}\, \zeta_2
-\, \frac{3}{4}\, L_e(x)\,
-\, \frac{1}{4}\, L_e^2(x)\, ,
\end{split}
\eq
%--
whereas $F^{(1)}_{\rm{M}}(x)$ vanishes when we neglect the electron mass,
$F^{(1)}_{\rm{M}}(x)\,=\, 0$.
The renormalized one-loop vertex develops an infrared divergency,
which shows up as a  single pole in the $\epsilon$ plane for $\epsilon=0$.
Therefore, when computing the cross section, we sum over the spins the squared matrix element 
and we evaluate the traces over Dirac indices in $D = 4\,-\,2\,\epsilon$ dimensions.
The needed kinematical structures include ${\cal O}(\epsilon)$ terms
(see Eq.~\eqref{kinCOEFF}).
%--
\subsubsection*{Irreducible Diagrams}
%--
The renormalized two-loop vertex diagrams of class \rm{2d}
are free of infrared divergencies.
Therefore, we can neglect  ${\cal O}(\epsilon)$ terms 
in the kinematical coefficients of Eq.~\eqref{kinCOEFF} 
appearing in Eq.~\eqref{VERTEX2}, setting $v_a(x,y,;\ep)=v_a(x,y;0)$,
for $a=1,2$.
%and write
%\bq
%\begin{split}
%\frac{d \sigma^{\rm{2d}\times\rm{tree}}}{d \Omega}
%\,=\,2\,
%\frac{\alpha^2}{s}\,
%\Bigl\{\,
%&\frac{1}{s^2}\,
%\Bigl[\,
%v_1(s,t;0)\,
%A_{\rm{V}}^{\rm{2d}}(s)\,
%\, +\,
%s^2\, 
%A_{\rm{M}}^{\rm{2d}}(s)
%\Bigr]
%+\,
%\frac{1}{t^2}\,
%\Bigl[\,
%v_1(t,s;0)\,
%A_{\rm{V}}^{\rm{2d}}(t)\,
%\, +\,
%t^2\, A_{\rm{M}}^{\rm{2d}}(t)\,
%\Bigr]\\
%+&  \frac{1}{s\, t}\, 
%\Bigl[\,
%v_2(s,t;0)\,
% \Bigl(\, A_{\rm{V}}^{\rm{2d}}(s)\,+\,
%A_{\rm{V}}^{\rm{2d}}(t)\,\Bigr)
%\,+\,
%\frac{3}{2}\,
%\Bigl(\,
%s^2\, A_{\rm{M}}^{\rm{2d}}(s)\,
%+\,
%t^2 \,A_{\rm{M}}^{\rm{2d}}(t)\,
%\Bigr)\,\\
%&\,\,\,\,\,\,\,+\,
%2\, s\,t\,
% \Bigl(\, A_{\rm{M}}^{\rm{2d}}(s)\,+\,
%A_{\rm{M}}^{\rm{2d}}(t)\,\Bigr)\,
%\Bigr]\,
%\Bigr\}.
%\end{split}
%\eq
%--
The auxiliary functions  $A_{\rm{V}}^{\rm{2d}}(x)$ and
$A_{\rm{M}}^{\rm{2d}}(x)$ contain the renormalized two-loop vector and magnetic vertex form factors 
(see~\cite{Bonciani:2003te,Mastrolia:2003yz,Bonciani:2003ai} for a detailed discussion),
\bq
A_{\rm{I}}^{\rm{2d}}(x)\, = \,
\sum_{f}\,
Q_f^2\,
\text{Re}\, \Bigl[\, F^{(2)}_{{\rm{I}},f}(x) \,  \Bigr],\qquad
\rm{I}\,=\rm{V,M}.
\eq
%--
For the case with an electron loop, $F^{(2)}_{\rm{I},e}(x)$, the exact results in terms
of Harmonic Polylogarithms, can be readily expanded in the high-energy limit.
For the vector term we get
%-
\bq\label{Vec2EL}
\begin{split}
F^{(2)}_{{\rm{V}},e}(x)\,&=\,
\frac{1}{4}\,\Bigl(\,
\frac{383}{27}\, -\, \zeta_2\, \Bigr)\, +\, \frac{1}{6}
\left( \, \frac{265}{36}\,+\, \zeta_2 \, \right)\,
L_e(x)\,+\,
\frac{19}{72}\, L_e^2(x)\,
+\,
\frac{1}{36}\, L_e^3(x).
\end{split}
\eq
%--
%%
%% Stefano: results corrected on 25 Jan 07, added reference to Burgers
%%
For $F^{(2)}_{\rm{V},f}(x)$, $f\neq e$, we perform an asymptotic expansion of the
Master Integrals arising in the computation (see Table V in~\cite{Czakon:2004wm}) 
and we fully agree with the result of \cite{Burgers:1985qg},
%--
\bq\label{Vec2MU}
\begin{split}
F^{(2)}_{{\rm{V}},f}(x)\,&=\,
\frac{1}{6}\,
\Bigl(\,
\frac{3355}{216}\,
+\, \frac{19}{6}\, \zeta_2\,
-\, 2\, \zeta_3\,
\Bigr)\,
+\, \frac{1}{6}
\left( \, \frac{265}{36}\,+\, \zeta_2 \, \right)\,
L_f(x)\,+\,
\frac{19}{72}\, L_f^2(x)\,
+\,
\frac{1}{36}\, L_f^3(x).
\end{split}
\eq
Since collinear logarithms are absent, the logarithmic
structure of Eqs.~\eqref{Vec2EL} and \eqref{Vec2MU} is obviously the same.
%--
\subsection{Box Corrections}\label{BOXES}
%--
The contribution of the renormalized two-loop box diagrams of class \rm{2e} is given by
\bq
\begin{split}
\frac{d \sigma^{\rm{2e}\times\rm{tree}}}{d \Omega}\,=\,
\frac{\alpha^2}{2\, s}\,
\Bigl[\,
\frac{1}{s}\, A_1^{\rm{2e}\times\rm{tree}}(s,t)\, +\,
\frac{1}{t}\, A_2^{\rm{2e}\times\rm{tree}}(s,t)
\,\Bigr]
.
\end{split}
\eq
%--
Here the auxiliary functions can be conveniently expressed through three
independent form factors $B^{(2)}_{\rm{I},f}(x,y)$, where $\rm{i}=\rm{A},\rm{B},\rm{C}$,
%--
\begin{eqnarray}
 A_1^{\rm{2e}\times\rm{tree}}(s,t)\, &=&\, F_\ep^2 \sum_{f}\,
Q_f^2\, \text{Re}\,
\Bigl[\, B^{(2)}_{{\rm{A}},f}(s,t)\, +\, B^{(2)}_{{\rm{B}},f}(t,s)\, + \, B^{(2)}_{{\rm{C}},f}(u,t)\,-\, B^{(2)}_{{\rm{B}},f}(u,s)\, \Bigr],
\\
 A_2^{\rm{2e}\times\rm{tree}}(s,t)\, &=&\,  F_\ep^2 \sum_{f}\,
Q_f^2\, \text{Re}\,
\Bigl[\, B^{(2)}_{{\rm{B}},f}(s,t)\, +\, B^{(2)}_{{\rm{A}},f}(t,s)\, - \, B^{(2)}_{{\rm{B}},f}(u,t)\,+\, B^{(2)}_{{\rm{C}},f}(u,s)\, \Bigr].
\end{eqnarray}
%--
\subsubsection*{Electron Loops}
For the case with an electron loop, $B^{(2)}_{\rm{I},e}(x,y)$, we get exact results in terms
of Harmonic Polylogarithms and Generalized Harmonic 
Polylogarithms. 
An asymptotic expansion in the limit $m_e^2\ll s,t,u$ leads to
%--
\allowdisplaybreaks
%-----------------------------------------------------
\begin{eqnarray}
\label{electronBOX2}
B_{{\rm{A}},e}^{(2)}(x,y)
&=&
\frac{1}{\epsilon}\,
\frac{2}{3}\,
\Bigl(\, \frac{x^2}{y}\, +\, 2\, x \, +\, y\, \Bigr)\,
\Bigl[\,
\frac{5}{3}\, +\,L_e(y)\,
\Bigr]\,
L_e(x)
+
\frac{1}{3}\,
\frac{x^2}{y}\,
\Bigl\{\,
-\, \frac{2}{3} \, \Bigl(\, \frac{17}{3}\,
+\,20 \, \zeta_2\, \Bigr)\, \nonumber \\
&+& 2\,
\Bigl(\,
\frac{41}{9}\,
-\, \zeta_2\,
\Bigr)\,
L_e(x)\,
-\,2\,
\Bigl(\,
\frac{1}{3}\,
+\, 8\, \zeta_2\,
\Bigr)\,
L_e(y)
- \frac{23}{6}\, L^2_e(y)
 \, +\, 8\, L_e(x)\, L_e(y)\nonumber\\
&-& \frac{5}{3}\, L^3_e(y)
+\, 4\, L_e(x)\, L^2_e(y)
- \Bigl[\,
6\, \zeta_2
\, +\, 
\ln^2\left(\frac{y}{x}\right)\,
%\Bigl(\,L_e(x)\,-\,L_e(y)\,\Bigr)^2\,
\Bigr]\,
\ln\left(1+\frac{y}{x}\right)
- 2\,
\ln \left(\frac{y}{x}\right)\,
%\Bigl[\,
%L_e(x)\,-\,L_e(y)\,
%\Bigr]\,
\text{Li}_{\rm{2}}\left(-\frac{y}{x}\right)\nonumber \\
&+& 2\,\text{Li}_{\rm{3}}\left(-\frac{y}{x}\right)\,
\Bigr\}
+ 
\frac{x}{3}\,
\Bigl\{\,
-\, \frac{2}{3}\, \Bigl(\, \frac{34}{3}\,
+\, 7\,  \zeta_2\, \Bigr)\,
+\,
\frac{242}{9}\, L_e(x)\,
-\, 4\, 
\Bigl(\,
\frac{5}{3}\,
+\, 6\, \zeta_2\,
\Bigr)\,
L_e(y) \nonumber \\
&+& \frac{1}{3} \, \Bigl[\,13 \,L^2_e(x)\,
 -\, 16\,L^2_e(y)
\, + \, 34 \,L_e(x)\,L_e(y)\, \Bigr]\,
  +\, 2\, \Bigl[\, \frac{1}{3}\, L_e^3(x)\,
-\, L^3_e(y)\, \nonumber\\
&+& 3\, L_e(x)\, L^2_e(y)\,
\Bigr]
 -  2\, \Bigl[\,
6 \,\zeta_2\,
+\, \ln^2\left(\frac{y}{x}\right)\,
%+\, \Bigl(\, L_e(x)\, -\, L_e(y)\, \Bigr)^2\,
\Bigr]\,
\ln\left(1+\frac{y}{x}\right) 
 -  4\,
\ln \left(\frac{y}{x}\right)\,
%\Bigl[\,
%L_e(x)\,-\, L_e(y)\,
%\Bigr]\,
\text{Li}_{\rm{2}}\left(-\frac{y}{x}\right)\nonumber\\
  &+& 4\, \text{Li}_{\rm{3}}\left(-\frac{y}{x}\right)\,
\Bigr\}
+ \frac{y}{3}\,
\Bigl\{\,
-\, \frac{2}{3}\, \Bigl(\, \frac{17}{3}\,
\,+\, 11\, \zeta_2\, \Bigr)\,
+\,
\frac{130}{9}\,
L_e(x)\,
-\, 6\,
\Bigl(\,
1\,
+\, 2\, \zeta_2\,
\Bigr)\,
L_e(y)\nonumber \\
& + &\frac{5}{3}\, \Bigl[ \,L^2_e(x)\,
 - \,\frac{5}{2}\, L^2_e(y)\,
+ \, 4 \, L_e(x)\,L_e(y)\,\Bigr]\,
+\, \frac{1}{3}\, L_e^3(x)\,
+\, 3\, L_e(x)\,L^2_e(y)\,
-\, L^3_e(y) \nonumber \\
& - & \Bigl[\,
6 \, \zeta_2\,
+\, 
\ln^2\left(\frac{y}{x}\right)
%\Bigl(\, L_e(x)\, -\, L_e(y)\, \Bigr)^2
\,
\Bigr]\,
\ln\left(1+\frac{y}{x}\right)
 -  2\,
\ln \left(\frac{y}{x}\right)\,
%\Bigl[\,
%L_e(x)\, -\, L_e(y)\,
%\Bigr]\,
\text{Li}_{\rm{2}}\left(-\frac{y}{x}\right)
\,   +\, 2\, \text{Li}_{\rm{3}}\left(-\frac{y}{x}\right)\,
\Bigr\}
,\\
&&\nonumber \\
%\end{eqnarray}
%--
%\begin{eqnarray}
B_{{\rm{B}},e}^{(2)}(x,y)
&=&
\frac{1}{\epsilon}\,
\frac{2}{3}\,
\Bigl(\, 
2\, \frac{x^2}{y}\, +\, 2\, x \, +\, y\, 
\Bigr)\,
\Bigl[\,
\frac{5}{3}\, +\,L_e(y)\,
\Bigr]\,
L_e(x)
+
\frac{1}{3}\,
\frac{x^2}{y}\,
\Bigl\{\,
\frac{4}{3}\,
\Bigl(\,
-\, \frac{17}{3}\,
-\, 20\, \zeta_2\,
\Bigr)\,\nonumber\\
&+& 4\,
\Bigl(\,
\frac{56}{9}\,
-\, \zeta_2\,
\Bigr)\, L_e(x)\,
-\, 4\, \Bigl(\,
\frac{1}{3}\,
+\, 8\, \zeta_2\,
\Bigr)\, L_e(y)
- \, \Bigl[\,
\frac{23}{3}\,L_e^2(y)\,
-\, 20\, L_e(x)\, L_e(y)\, \Bigr]\,\nonumber\\
&-&2 \,\Bigl[\,
\frac{5}{3}\,L_e^3(y)\,
-\,4\,L_e(x)\,L_e^2(y)\,\Bigr]
- 2\, \Bigl[\,
6 \,\zeta_2\,
+\, \ln^2\left(\frac{y}{x}\right)\,
%+\,\Bigl(\, L_e(x)\,-\,L_e(y)\, \Bigr)^2\,
\Bigr]\, \ln\left(1+\frac{y}{x}\right)\nonumber\\
&-& 4\,
\ln \left(\frac{y}{x}\right)
%\Bigl[
%\,L_e(x)\,-\,L_e(y)\,
%\Bigr]
\,\text{Li}_2\left(-\frac{y}{x}\right)\,
+ 4\, \text{Li}_3\left(-\frac{y}{x}\right)\,
\Bigr\}
+ \frac{x}{3}\,
\Bigl\{\,
-\,\frac{2}{3}\,\Bigl(\,
\frac{34}{3}\,
+\,7\,\zeta_2\,
\Bigr)\,
+\,
\frac{272}{9}\,L_e(x)\,\nonumber\\
&-& 4\, \Bigl(\,
\frac{5}{3}\,
+\,6\, \zeta_2\,
\Bigr)\, L_e(y)
+
\frac{1}{3}\, \Bigl[\,
13\, L_e^2(x)\,
+\,40\, L_e(x)\,L_e(y)\, - 16\, L_e^2(y)\, \Bigr]\,\nonumber\\
&+& 2\, \Bigl[\, \frac{1}{3}\, L_e^3(x)\,
-\, L_e^3(y)\,
+\,3\, L_e(x)\,L_e^2(y)\,\Bigr]
- 2\, \Bigl[\,
6 \,\zeta_2\,
+\, \ln^2\left(\frac{y}{x}\right)\,
%+\, \Bigl(\, L_e(x)\,-\,L_e(y)\,\Bigr)^2\,
\Bigr]\, \ln\left(1+\frac{y}{x}\right)\nonumber\\
&-& 4\,
 \ln \left(\frac{y}{x}\right)
%\Bigl[\,
%L_e(x)\,-\,L_e(y)\,
%\Bigr]\, 
\text{Li}_2\left(-\frac{y}{x}\right)\,
+\,4\,\text{Li}_3\left(-\frac{y}{x}\right)\,
\Bigr\}
+ 
\frac{y}{3}\,
\Bigl\{\,
-\,\frac{2}{3}\, \Bigl(\,
\frac{17}{3}\,
+\, 11\, \zeta_2\,\Bigr)\,
+\,
\frac{130}{9}\,L_e(x)\,\nonumber\\
&-& 6\, \Bigl(\,
1\,
+\,2\, \zeta_2\,
\Bigr)\, L_e(y)
+\frac{5}{3}\,
\Bigl[\,
L_e^2(x)\,
-\,\frac{5}{2}\, L_e^2(y)\,
 +\,4\, L_e(x)L_e(y)\,\Bigr]\,
+\,\frac{1}{3}\, L_e^3(x)\,\nonumber\\
&-&  L_e^3(y)\,
+\,3\, L_e(x)\,L_e^2(y)
- \Bigl[\,
6\, \zeta_2\,
+\, \ln^2\left(\frac{y}{x}\right)\,
%+\,\Bigl(\, L_e(x)\,-\,L_e(y)\,\Bigr)^2\,
\Bigr]\, \ln\left(1+\frac{y}{x}\right)\nonumber\\
&-&  2\,
 \ln\left(\frac{y}{x}\right)\,
%\Bigl[\,
%L_e(x)\,-\,L_e(y)\,
%\Bigr]\, 
\text{Li}_2\left(-\frac{y}{x}\right)\,
+ 2\, \text{Li}_3\left(-\frac{y}{x}\right)\,
\Bigr\}
,\\
&&\nonumber \\
%\end{eqnarray}
%---
%\begin{eqnarray}
B_{{\rm{C}},e}^{(2)}(x,y)
&=&
- \, \frac{1}{\epsilon}\,
\frac{2}{3}\,
\frac{x^2}{y}\,
\Bigl[\,
\frac{5}{3}\, +\,L_e(y)\,
\Bigr]\,
L_e(x)\,
+\, \frac{2}{3}\,\Bigl(\,x\,+\,y\,\Bigr)\,\,
\Bigl[\,
\frac{5}{3}\,
+\,L_e(y)\,
\Bigr]\,
L_e(x)
\nonumber \\
&+&\frac{1}{3}\,
\frac{x^2}{y}\,
\Bigl\{\,
\frac{2}{3}\,
\Bigl(\,
\frac{17}{3}\,
+\,20\,\zeta_2\,\Bigr)\,
-\, 2\,
\Bigl(\,
\frac{41}{9}\,
-\,\zeta_2\,\Bigr)\,
L_e(x)\,
+\,2\,
\Bigl(\,
\frac{1}{3}\,
+\,8\,\zeta_2\,\Bigr)\,
L_e(y)\nonumber \\
& 
+&\frac{23}{6}\, L_e^2(y)\,
-\,8\, L_e(x)\, L_e(y)\,
+\,\frac{5}{3}\, L_e^3(y)\,
-\, 4\, L_e(x)\, L_e^2(y)\nonumber \\
&+&
\Bigl[\,
6\, \zeta_2\,
+\, \ln^2\left(\frac{y}{x}\right)\,
%+\,\Bigl(\, L_e(x)\,-\,L_e(y)\,\Bigr)^2\,
\Bigr]\,
\ln\left(1+\frac{y}{x}\right)
+ 2\,
 \ln\left(\frac{y}{x}\right)\,
%\Bigl[\,
%L_e(x)\,-\,L_e(y)\,
%\Bigr]\,
\text{Li}_2\left(-\frac{y}{x}\right)\,
-\, 2\,  \text{Li}_3\left(-\frac{y}{x}\right)
\, \Bigr\}
.
\end{eqnarray}
%--
\subsubsection*{Heavy-Fermion Loops}
The list of Master Integrals here is given in Table V of~\cite{Czakon:2004wm}).
At variance with the electron-loop case, it is not
possible to compute them exactly by means of a basis containing Harmonic
Polylogarithms and Generalized Harmonic Polylogarithms.
Therefore, we use the high-energy asymptotic expansion discussed
in Subsection~\ref{outline}.
The results, expressed by the logarithms of the fermion
masses $L(R_f)$ (see Eq.~\eqref{LOG2}), are:
%%
%% Stefano: results added on 25 Jan 07
%%
\begin{eqnarray}
\label{muonBOX2}
B_{{\rm{A}},f}^{(2)}(x,y)
&=& 
\frac{1}{\epsilon}\,
\frac{2}{3}\,
\Bigl(\, \frac{x^2}{y}\, +\, 2\, x \, +\, y\, \Bigr)\,
\Bigl[\,
\frac{5}{3}\, -\,L(R_f)\, +\, L_e(y)\,
\Bigr]\,
L_e(x)
\nonumber \\
&+&
\frac{1}{3}\,
\frac{x^2}{y}\,
\Bigl\{\,
2\,
\Bigl(\,
\frac{131}{27}\,
-\, 10\, \zeta_2\,
-\, 2\, \zeta_3
\, \Bigr)\,
-\,2\, \Bigl(\,\frac{25}{9}\,-\, 6\, \zeta_2 \, \Bigr)\, L(R_f)\,
+\, \frac{7}{6}\, L^2(R_f)\,\nonumber\\
&-& \frac{1}{3}\, L^3(R_f)
+ \Bigl[\,
\frac{82}{9}\, -\, 2\, \zeta_2\,
-\, \frac{4}{3}\, L(R_f)
\,\Bigr]\, L_e(x)\,
-\,2\,
\Bigl[\,
\frac{1}{3}\, +\, 8\, \zeta_2\, -\,\frac{1}{2}\, L(R_f)\,
\Bigr]
\, L_e(y)\nonumber \\
& - & \Bigl[\,\frac{23}{6}\,
-\,2\, L(R_f)\, \Bigr]\, L^2_e(y)\,
+\, 4\, 
\Bigl[\,
2\, -\, L(R_f)
\,\Bigr]\,L_e(x)\, L_e(y)\,
-\, 4\, \Bigl[\,
\frac{5}{12}\, L_e^3(y)\,\nonumber\\
&-& L_e(x)\, L_e^2(y)
\, \Bigr]
- \Bigl[\,
6\, \zeta_2\, +\,
\ln^2\left(\frac{y}{x}\right)
%\Bigl(\,
%L_e(x)\,-\,L_e(y)
%\,\Bigr)^2
\,\Bigr]
\ln\left(1+\frac{y}{x}\right)
- 2\,
\ln \left(\frac{y}{x}\right)
%\Bigl(\,
%L_e(x)\, -\, L_e(y)
%\,\Bigr)
\text{Li}_{\rm{2}}\left(-\frac{y}{x}\right)\nonumber\\
&+& 2\,
\text{Li}_{\rm{3}}\left(-\frac{y}{x}\right)
\,\Bigr\}
+ 
\frac{x}{3}\,
\Bigl\{\,
2\,\Bigl(\, \frac{262}{27}\,-\,9\,\zeta_2\,-4\,\zeta_3\,\Bigr)\,
-\,4\,\Bigl(\,\frac{25}{9}\,-\,3\,\zeta_2\,\Bigr)\,
L(R_f)\,
+\,\frac{7}{3}\,L^2(R_f)\nonumber\\
&-&\frac{2}{3}\, L^3(R_f)
 + 2\,
\Bigl[\,
\frac{121}{9}\,-\,\frac{10}{3}\,L(R_f)\,
\Bigr]\,L_e(x)\,
-\,2\,
\Bigl[\,
\frac{10}{3}\,+\,12\, \zeta_2\,-\,2\,L(R_f)\,
\Bigr]\,L_e(y)\nonumber \\
&+ & \Bigl[\, \frac{13}{3}\,-\,2\,L(R_f)\,\Bigr]\,L_e^2(x)\,
-\,\Bigl[\, \frac{16}{3}\,-\,2\,L(R_f)\,\Bigr]\,L_e^2(y)
 + 2\,\Bigl[\,\frac{17}{3}\,-\,2\,L(R_f)\, \Bigr]\,L_e(x)\,L_e(y)\,\nonumber\\
&+&\frac{2}{3}\,L_e^3(x)\,
+\,6\,L_e(x)\,L^2_e(y)\,
-\,2\,L_e^3(y)
 - 2 \Bigl[\,
6\,\zeta_2\,+
\, \ln^2\left(\frac{y}{x}\right)\,
%\,\Bigl(\,L_e(x)\,-\,L_e(y)\,\Bigr)^2\,
\Bigr]\,\ln\left(1+\frac{y}{x}\right)\nonumber\\
& -& 4\,
\ln \left(\frac{y}{x}\right)
%\Bigl( \,L_e(x)\,-\,L_e(y)\,\Bigr)
\,
\text{Li}_2\left(-\frac{y}{x}\right)\,
+\,4\,\text{Li}_3\left(-\frac{y}{x}\right)\,
\Bigr\}
+ \frac{y}{3}\,
\Bigl\{\,
2\Bigl(\frac{131}{27}-7\zeta_2-2\zeta_3 \Bigr)\nonumber\\
&-&2\Bigl(\frac{25}{9}-3\zeta_2 \Bigr)L(R_f)
+\frac{7}{6}L^2(R_f)-\frac{1}{3}L^3(R_f)
 + \Bigl[\frac{130}{9}-\frac{10}{3}L(R_f)\Bigr]L_e(x)\nonumber\\
&-&\Bigl[ 6+12\zeta_2-3L(R_f)\Bigr]L_e(y)
+\Bigl[\frac{5}{3}-L(R_f)\Bigr]L_e^2(x)
-\Bigl[\frac{25}{6}-L(R_f)\Bigr]L_e^2(y)\nonumber\\
&+&2\Bigl[\frac{10}{3}-L(R_f)\Bigr]L_e(x)L_e(y)
 +\frac{1}{3} L_e^3(x) - L_e^3(y)+3 L_e(x)L_e^2(y)\nonumber\\
& -&\Bigl[
6\,\zeta_2\,+\,
\ln^2\left(\frac{y}{x}\right)\,
%\Bigl(\,L_e(x)\,-\,L_e(y)\,\Bigr)^2\,
\Bigr]\,\ln\left(1+\frac{y}{x}\right)
 - 2
\ln\left(\frac{y}{x}\right)\,
%\Bigl( \,L_e(x)\,-\,L_e(y)\,\Bigr)\,
\text{Li}_2\left(-\frac{y}{x}\right)\,
+2 \, \text{Li}_3\left(-\frac{y}{x}\right)\,
\Bigr\},\\
&&\nonumber \\ \label{muonBOX3}
%\end{eqnarray}
%
%\begin{eqnarray}
B_{{\rm{B}},f}^{(2)}(x,y)
&=&
\frac{1}{\epsilon}\,
\frac{2}{3}\,
\Bigl(\, 2\, \frac{x^2}{y}\, +\, 2\, x \, +\, y\, \Bigr)\,
\Bigl[\,
\frac{5}{3}\, -\,L(R_f)\, +\, L_e(y)\,
\Bigr]\,
L_e(x)
\nonumber \\
&+&
\frac{2}{3}\,
\frac{x^2}{y}\,
\Bigl\{\,
\frac{262}{27}\,
-\, 20\, \zeta_2\,
-\, 4\, \zeta_3
-\, \Bigl(\,\frac{50}{9}\,-\, 12\, \zeta_2 \, \Bigr)\, L(R_f)\,
+\, \frac{7}{6}\, L^2(R_f)\,
-\, \frac{1}{3}\, L^3(R_f)\nonumber \\
& + & \Bigl[\,
\frac{112}{9}\, -\, 2\, \zeta_2\,
-\, \frac{10}{3}\, L(R_f)
\,\Bigr]\, L_e(x)\,
+\,
\Bigl[\,
-\, \frac{2}{3}\, -\, 16\, \zeta_2\, + \, L(R_f)\,
\Bigr]
\, L_e(y)\nonumber \\
& - & \Bigl[\,\frac{23}{6}\,
-\,2\, L(R_f)\, \Bigr]\, L^2_e(y)\,
+\, 2\, 
\Bigl[\,
5\, -\, 2\, L(R_f)
\,\Bigr]\,L_e(x)\, L_e(y)\,
-\, 4\, \Bigl[\,
\frac{5}{12}\, L_e^3(y)\,\nonumber\\
&-& L_e(x)\, L_e^2(y)
\, \Bigr]
-  \Bigl[\,
6\, \zeta_2\, +\,
\ln^2\left(\frac{y}{x}\right)
%\Bigl(\,
%L_e(x)\,-\,L_e(y)
%\,\Bigr)^2
\,\Bigr]
\ln\left(1+\frac{y}{x}\right)
-  2\,
\ln \left(\frac{y}{x}\right)\,
%\Bigl(\,
%L_e(x)\, -\, L_e(y)
%\,\Bigr)
\text{Li}_{\rm{2}}\left(-\frac{y}{x}\right)\,
+\, 2\,
\text{Li}_{\rm{3}}\left(-\frac{y}{x}\right)
\,\Bigr\} \nonumber \\
&+& 
\frac{2\, x}{3}\,
\Bigl\{\,
\frac{262}{27}\,-\,9\,\zeta_2\,-4\,\zeta_3\,
-\,2\,\Bigl(\,\frac{25}{9}\,-\,3\,\zeta_2\,\Bigr)\,
L(R_f)\,
+\,\frac{7}{6}\,L^2(R_f)\,
-\,\frac{1}{3}\, L^3(R_f)\nonumber \\
& + &
\Bigl[\,
\frac{136}{9}\,-\,\frac{13}{3}\,L(R_f)\,
\Bigr]\,L_e(x)\,
-\,
\Bigl[\,
\frac{10}{3}\,+\,12\,\zeta_2\,-\,2\,L(R_f)\,
\Bigr]\,L_e(y)\nonumber \\
& + & \Bigl[\, \frac{13}{6}\,-\,L(R_f)\,\Bigr]\,L_e^2(x)\,
-\,\Bigl[\, \frac{8}{3}\,-\,L(R_f)\,\Bigr]\,L_e^2(y)
 + \Bigl[\,\frac{20}{3}\,-\,2\,L(R_f)\, \Bigr]\,L_e(x)\,L_e(y)\,\nonumber\\
&+& \frac{1}{3}\,L_e^3(x)\,
+\,3\,L_e(x)\,L^2_e(y)\,
-\,L_e^3(y)
 - \Bigl[\,
6\,\zeta_2\
+ 
\ln^2\left(\frac{y}{x}\right)\,
%\Bigl(\,L_e(x)\,-\,L_e(y)\,\Bigr)^2\,
\Bigr]\,\ln\left(1+\frac{y}{x}\right)\nonumber\\
 &-& 2 
\ln \left(\frac{y}{x}\right)\,
%\Bigl( \,L_e(x)\,-\,L_e(y)\,\Bigr)
\,\text{Li}_2\left(-\frac{y}{x}\right)\,
+2\,\text{Li}_3\left(-\frac{y}{x}\right)\,
\Bigr\}
+ \frac{2\, y}{3}\,
\Bigl\{\,
\Bigl(\frac{131}{27}-7\zeta_2 \nonumber\\
&-&2\zeta_3 \Bigr)
-\, \Bigl(\frac{25}{9}-3\zeta_2 \Bigr)L(R_f)
+\frac{7}{12}L^2(R_f)-\frac{1}{6}L^3(R_f)
 +\Bigl[\frac{65}{9}-\frac{5}{3}L(R_f)\Bigr]L_e(x)\nonumber\\
&-&\frac{1}{2}\, \Bigl[ 6+12\zeta_2-3L(R_f)\Bigr]L_e(y)
 +\frac{1}{2}\, \Bigl[\frac{5}{3}-L(R_f)\Bigr]L_e^2(x)
-\, \frac{1}{2}\, \Bigl[\frac{25}{6}-L(R_f)\Bigr]L_e^2(y)\nonumber\\
&+&\Bigl[\frac{10}{3}-L(R_f)\Bigr]L_e(x)L_e(y)
 +\frac{1}{6}L_e^3(x)-\frac{1}{2}L_e^3(y)+\frac{3}{2}L_e(x)L_e^2(y)\nonumber\\
 &-&\Bigl[
3\,\zeta_2\,+\,\frac{1}{2}\, 
\ln^2\left(\frac{y}{x}\right)
%\Bigl(\,L_e(x)\,-\,L_e(y)\,\Bigr)^2
\,
\Bigr]\,\ln\left(1+\frac{y}{x}\right)
 - 
\ln\left(\frac{y}{x}\right)
%\Bigl( \,L_e(x)\,-\,L_e(y)\,\Bigr)
\,\text{Li}_2\left(-\frac{y}{x}\right)\,
+\,  \text{Li}_3\left(-\frac{y}{x}\right)\,
\Bigr\},\\
&&\nonumber \\\label{muonBOX4}
%\end{eqnarray}
%--
%\begin{eqnarray}
B_{{\rm{C}},f}^{(2)}(x,y)
&=& -\,
\frac{1}{\epsilon}\,
\frac{2}{3}\,
\frac{x^2}{y}\,
\Bigl[\,
\frac{5}{3}\, -\,L(R_f)\, +\, L_e(y)\,
\Bigr]\,
L_e(x)
 +
\frac{2}{3}\,
\Bigl( x + y \Bigr)
\Bigl[\,
\frac{5}{3}\, -\,L(R_f)\, +\, L_e(y)\,
\Bigr]\,
L_e(x)\nonumber \\
& +&
\frac{2}{3}\,
\frac{x^2}{y}\,
\Bigl\{\,
-\frac{131}{27}
+ 10 \,\zeta_2
+ 2 \,\zeta_3
+ \Bigl(
\frac{25}{9} - 6\, \zeta_2
\Bigr) L(R_f)
- \frac{7}{12} L^2(R_f)
+ \frac{1}{6} L^3(R_f)\nonumber \\
&
-& \Bigl(
\frac{41}{9}
- \zeta_2
-\frac{2}{3} L(R_f)
\Bigr)\, L_e(x)\,
+\, \Bigl(
\frac{1}{3}
+\, 8\, \zeta_2
-\frac{1}{2} L(R_f)
\Bigr)\, L_e(y)\nonumber \\
&
-& 2\ \Bigl(
2 - \, L(R_f)
\Bigr) L_e(x)\, L_e(y)
+ \Bigl(
\frac{23}{12} - L(R_f)
\Bigr) L_e^2(y)
+ \frac{5}{6} L_e^3(y)
- 2 L_e(x) L_e^2(y)\nonumber\\
&+& \Bigl[
3\,\zeta_2\,+\,\frac{1}{2}\, 
\ln^2\left(\frac{y}{x}\right)
%\Bigl(\,L_e(x)\,-\,L_e(y)\,\Bigr)^2
\,
\Bigr]\,\ln\left(1+\frac{y}{x}\right)
 +  
\ln \left(\frac{y}{x}\right)
%\Bigl( \,L_e(x)\,-\,L_e(y)\,\Bigr)
\,\text{Li}_2\left(-\frac{y}{x}\right)\,
-\, \text{Li}_3\left(-\frac{y}{x}\right)\,
\, \Bigr\}
.
\end{eqnarray}
In order to study the numerical effects of massive leptons in two-loop box diagrams
we consider the interference of the box diagram of class {\rm 2e}
(see Figure~\ref{2loop}) with the s-channel tree-level amplitude,
\begin{equation}\label{boxNUMS}
B_{{\rm 2e},f}\, =\,
\frac{\alpha^2}{4\, s^2} 
\text{Re} \Bigl[
B_{A,f}^{(2)}(s,t)
\Bigr],
\end{equation}
where $B_{A,f}$ can be found in Eq.~\eqref{electronBOX2} for electron loops,
and in Eq.~\eqref{muonBOX2} for $f\neq e$ loops.
In Table~\ref{box3} (Table~\ref{box90}) we show numerical values for the finite part of
$B_{{\rm 2e},f}$ at values of $\sqrt{s}$ typical for meson factories, Giga-Z, ILC, and at two selected small and wide scattering angles, $\theta=3^\circ$ ($\theta=90^\circ$).
\begin{table}[ht]\centering
\setlength{\arraycolsep}{\tabcolsep}
\renewcommand\arraystretch{1.2}
\begin{tabular}{|l|c|l|l|}
\hline 
$B_{{\rm 2e},f}$ [nb] $\slash$ $\sqrt{s}$\ [GeV]    & 10             & 91   & 500 \\
\hline 
\hline
$e$ [see Eq.~\eqref{electronBOX2}] &  188758         & 5200.08 & 284.711 \\
\hline 
$\mu$ [see Eq.~\eqref{muonBOX2}] & 1635.62        &1686.88  & 130.579  \\
\hline 
$\tau$ \qquad `` &           &  & 39.5554  \\
\hline 
\end{tabular}
\caption[]{Numerical values for the finite part of $B_{{\rm 2e},f}$ of Eq.~\eqref{boxNUMS}
in nanobarns at a scattering angle $\theta=3^\circ$.
The first two entries for the $\tau$ lepton are not shown since here the high-energy
approximation in not justified (the same consideration applies to the top quark).}
\label{box3}
\end{table}
\begin{table}[ht]\centering
\setlength{\arraycolsep}{\tabcolsep}
\renewcommand\arraystretch{1.2}
\begin{tabular}{|l|c|l|l|}
\hline 
$B_{{\rm 2e},f}$ [nb] $\slash$ $\sqrt{s}$\ [GeV]    & 10             & 91   & 500 \\
\hline 
\hline
$e$ [see Eq.~\eqref{electronBOX2}] &  143.162       & 3.23102 & 0.160582 \\
\hline 
$\mu$ [see Eq.~\eqref{muonBOX2}] & 61.3875       & 1.79381 & 0.0995184  \\
\hline 
$\tau$ \qquad `` & 10.0105         & 0.935319 & 0.0639576 \\
\hline 
t \qquad `` &          &  & -0.00256757 \\
\hline 
\end{tabular}
\caption[]{Numerical values for the finite part of $B_{{\rm 2e},f}$ of Eq.~\eqref{boxNUMS}
in nanobarns at a scattering angle $\theta=90^\circ$.
The first two entries for the top quark are not shown since here the high-energy
approximation in not justified.}
\label{box90}
\end{table}

For comparison we show in Figure \ref{vertex} the real part of the vertex function, see Eq.~\eqref{Vec2MU}.

\begin{table}[ht]\centering
\setlength{\arraycolsep}{\tabcolsep}
\renewcommand\arraystretch{1.2}
\begin{tabular}{|l|c|l|l|}
\hline 
$\sqrt{s}$\ [GeV]    & 10             & 91   & 500 \\
\hline 
\hline
$e$ &  -124.237              & -254.293   & -400.574 \\
\hline 
$\mu$ & -4.8036             & -29.1057  & -70.1032   \\
\hline 
$\tau$ &          & -2.08719 & -13.4901  \\
\hline 
\end{tabular}
\caption[]{The real part for the vertex form factor, see Eqs.~\eqref{Vec2EL} and \eqref{Vec2MU}.}
\label{vertex}
\end{table}

We see that the contributions from the box diagrams with heavier fermions are not strongly suppressed,  but are instead of about the same size as the boxes with electron loop.
This is different to the self-energy and  vertex corrections and may be traced back to the logarithmic structure of the contributions
Eqs.~\eqref{muonBOX2}--\eqref{muonBOX4}, where terms of the order $L_e^3(x)$ appear.
Further, in Eq.~\eqref{b5l2m2m} we may see that this Master Integral has a dependence on $L_e^3(x)$,  in contrast to the vertex and self-energy masters with heavy fermion loops.
That originates in an additional collinear mass singularity from the external legs of this diagram diagram.
One may control this easily by evaluating the singularity structure of the corresponding massless box diagram where only a scale $M$ due to the internal loop exists, and see there some $1/\ep^2$ terms which are absent in the corresponding SE and vertex  diagrams.
This leads finally to the fact that the two-loop corrections from heavier fermions are not numerically suppressed compared to the electron loop contributions.

%%%%%%%%%%%%%%%%%%%%%%%%%%%%%%%%%%%%%%%%%%%%%%%%%%%%%%%%%%%%%%%%%%%%%%%%%%%%%%%%%
\subsection{Products of One-Loop Corrections}
%--
Finally, we consider the simpler components generated by the interference of one-loop diagrams
among themselves.
We start with the interference of diagrams of class \rm{1a},
%--
\bq\label{loopBYloop}
\begin{split}
\frac{d \sigma^{\rm{1a}\times\rm{1a}}}{d \Omega}\,=\,
\frac{\alpha^2}{2\, s}\,
\Bigl\{\,
&\frac{1}{s^2}\,
v_1(s,t;0)\,
A^{\rm{1a}\times\rm{1a}}(s,s)\,
+\, \frac{1}{t^2}\,
v_1(t,s;0)\,
A^{\rm{1a}\times\rm{1a}}(t,t) \\
+& \frac{1}{s\, t}\,
v_2(s,t;0)\,
\Bigl[\,
 A^{\rm{1a}\times\rm{1a}}(s,t)
\, +\,
 A^{\rm{1a}\times\rm{1a}}(t,s)
\, \Bigr]
\,
\Bigr\}.
\end{split}
\eq
%--
Here the auxiliary function $A^{\rm{1a}\times\rm{1a}}(x,y)$
contains the product of the renormalized one-loop
vacuum-polarization function $\Pi^{(1)}_f(x)$ (see Eq.~\eqref{Pi10})  with its
complex conjugate,
%--
\bq
A^{\rm{1a}\times\rm{1a}}(x,y)\,\equiv\,
 \sum_{f_1,f_2} \, Q_{f_1}^2\, Q_{f_2}^2\, 
\Pi^{(1)}_{f_1}(x)\, \left[\Pi^{(1)}_{f_2}(y)\right]^\star.
\eq
%--
The interference of diagrams of class \rm{1a} with
those of class  \rm{1b} gives
%--
\begin{eqnarray}
\frac{d \sigma^{\rm{1a}\times\rm{1b}}}{d \Omega}\,=\,2\,
\frac{\alpha^2}{s}
&\Bigl\{&
\frac{1}{s^2}\,
\Bigl[\,
v_1(s,t;\epsilon)\,
A_{\rm{V}}^{\rm{1a}\times\rm{1b}}(s,s)\,
\, +\,
s^2\, 
A_{\rm{M}}^{\rm{1a}\times\rm{1b}}(s,s)
\Bigr] \nonumber \\
&+&
\frac{1}{t^2}\,
\Bigl[\,
v_1(t,s;\epsilon)\,
A_{\rm{V}}^{\rm{1a}\times\rm{1b}}(t,t)\,
\, +\,
t^2\, A_{\rm{M}}^{\rm{1a}\times\rm{1b}}(t,t)\,
\Bigr]\nonumber \\
&+&  \frac{1}{s\, t}\, 
\Bigl[\,
v_2(s,t;\epsilon)\,
 \Bigl(\, A_{\rm{V}}^{\rm{1a}\times\rm{1b}}(s,t)\,+\,
A_{\rm{V}}^{\rm{1a}\times\rm{1b}}(t,s)\,\Bigr) \nonumber \\
&+&
\frac{3}{2}\,
\Bigl(\,
s^2\, A_{\rm{M}}^{\rm{1a}\times\rm{1b}}(s,t)\,
+\,
t^2 \,A_{\rm{M}}^{\rm{1a}\times\rm{1b}}(t,s)\,
\Bigr)\,\nonumber\\
&+&
2\, s\,t\,
 \Bigl(\, A_{\rm{M}}^{\rm{1a}\times\rm{1b}}(s,t)\,+\,
A_{\rm{M}}^{\rm{1a}\times\rm{1b}}(t,s)\,\Bigr)\,
\Bigr]\,
\Bigr\}.
\end{eqnarray}
%--
The auxiliary function $A^{\rm{1a}\times\rm{1b}}(x,y)$ is given
by the product of $F^{(1)}_{\rm{V}}(x)$ and $F^{(1)}_{\rm{M}}(x)$, the renormalized one-loop 
vector (see Eq.~\eqref{Vec1}) and magnetic (vanishing in the high-energy limit) form factors
for the QED vertex, and the complex-conjugate renormalized one-loop
vacuum-polarization function $\Pi^{(1)}_f(x)$ (see Eq.~\eqref{Pi10}),
%--
\bq
A_{\rm{I}}^{\rm{1a}\times\rm{1b}}(x,y)\, \equiv \,
\sum_{f}\,
Q_f^2\,
\text{Re}\, \Bigl\{\, F^{(1)}_{\rm{I}}(x) \, \left[\Pi^{(1)}_f(y)\right]^\star\, \Bigr\},\qquad
\rm{I}\,=\rm{V,M}.
\eq
%--
Finally, the interference of diagrams of class \rm{1a} with
those of class  \rm{1c} gives
\bq
\begin{split}
\frac{d \sigma^{\rm{1a}\times\rm{1c}}}{d \Omega}\,=\,
\frac{\alpha^2}{4\, s}
\Bigl[
\frac{1}{s} \, A_1^{\rm{1a}\times\rm{1c}}(s,t)\, +\,
\frac{1}{t} \, A_2^{\rm{1a}\times\rm{1c}}(s,t)\,
\Bigr]
.
\end{split}
\eq
Here the auxiliary functions $A_1^{\rm{1a}\times\rm{1c}}(s,t)$ and 
$A_2^{\rm{1a}\times\rm{1c}}(s,t)$ take the form
%--
\bq
\begin{split}
 A_1^{\rm{1a}\times\rm{1c}}(s,t)\, =&\, F_\ep \sum_{f}\,
Q_f^2\, \text{Re}\, \left\{
\Bigl[\, B^{(1)}_{\rm{A}}(s,t)\, +\, B^{(1)}_{\rm{B}}(t,s)\, + \, B^{(1)}_{\rm{C}}(u,t)\,-\, B^{(1)}_{\rm{B}}(u,s)\, \Bigr]
\,  \left[\Pi^{(1)}_f(s)\right]^\star\right\},
\\
 A_2^{\rm{1a}\times\rm{1c}}(s,t)\, =&\,  F_\ep \sum_{f}\,
Q_f^2\, \text{Re}\, \left\{
\Bigl[\, B^{(1)}_{\rm{B}}(s,t)\, +\, B^{(1)}_{\rm{A}}(t,s)\, - \, B^{(1)}_{\rm{B}}(u,t)\,+\, B^{(1)}_{\rm{C}}(u,s)\, \Bigr]
\,  \left[\Pi^{(1)}_f(t)\right]^\star\right\}.
\end{split}
\eq

\begin{eqnarray}
 A_1^{\rm{1a}\times\rm{1c}}(s,t)\, &=&\, F_\ep \sum_{f}\,
Q_f^2\, \text{Re}\, \left\{
\Bigl[\, B^{(1)}_{\rm{A}}(s,t)\, +\, B^{(1)}_{\rm{B}}(t,s)\, + \, B^{(1)}_{\rm{C}}(u,t)\,-\, B^{(1)}_{\rm{B}}(u,s)\, \Bigr]
\,  \left[\Pi^{(1)}_f(s)\right]^\star\right\},
\nonumber
\\
\\
 A_2^{\rm{1a}\times\rm{1c}}(s,t)\, &=&\,  F_\ep \sum_{f}\,
Q_f^2\, \text{Re}\, \left\{
\Bigl[\, B^{(1)}_{\rm{B}}(s,t)\, +\, B^{(1)}_{\rm{A}}(t,s)\, - \, B^{(1)}_{\rm{B}}(u,t)\,+\, B^{(1)}_{\rm{C}}(u,s)\, \Bigr]
\,  \left[\Pi^{(1)}_f(t)\right]^\star\right\}.
\nonumber
\\
\end{eqnarray}

$\Pi^{(1)}_f(x)$ is given in Eq.~\eqref{Pi10}, and the new functions, in the small mass limit,  read as
%--
%\bq
\begin{eqnarray}
B_{\rm{A}}^{(1)}(x,y)
&=& 
-\, \frac{4}{\epsilon}\,
\Bigl(\, \frac{x^2}{y}\,+\,2\,x\,+\,y\,\Bigr)\, L_e(x)
+
\frac{x^2}{y}\,
\Bigl[\,
16\, \zeta_2\,
+\, 4\, L_e(x)\,
+ 2\, L_e^2(y)\,\nonumber\\
&-& 4\, L_e(x)\, L_e(y)\,
\, \Bigr]
+ 2\, x\,
\Bigl[\,
10\, \zeta_2\,
+\, L_e(x)\,
+\, L_e(y)\,
-\, L_e^2(x)\,
+ L_e^2(y)\,\nonumber\\
&-& 2\, L_e(x)\, L_e(y)\,
\,\Bigr]
+ y\,
\Bigl[\,
 10\, \zeta_2
+\, 2\, L_e(x)\,
+\, 2\, L_e(y)\,
- L_e^2(x)\,
+ L_e^2(y)\,\nonumber\\
&-& 2\, L_e(x)\, L_e(y)\,
\Bigr]
,\\
&&\nonumber \\
%\end{split}
%\eq
%--
%\bq
%\begin{split}
B_{\rm{B}}^{(1)}(x,y)
&=& 
-\, \frac{4}{\epsilon}\,
\Bigl(\, 2\, \frac{x^2}{y}\,+\,2\,x\,+\,y\,\Bigr)\, L_e(x)
+  4\, \frac{x^2}{y}\,
\Bigl[\,
 8\, \zeta_2\,
+\, L^2_e(y)\,
- 2\, L_e(x)\, L_e(y)\,
\, \Bigr]\nonumber\\
&+& 2\, x\,
\Bigl[\,
 10\, \zeta_2
-\,  L_e(x)\,
+\,  L_e(y)\,
-  L^2_e(x)\,
+\,  L^2_e(y)\,
-\, 2\, L_e(x)\, L_e(y)
\,\Bigr]\nonumber\\
&+& y\,
\Bigl[\,
10\, \zeta_2
+\, 2\, L_e(x)\,
+\, 2\, L_e(y)\,
- L^2_e(x)\,
+\, L^2_e(y)\,
-\, 2\, L_e(x)\, L_e(y)\,
\Bigr]
,\\
&& \nonumber\\
%\end{split}
%\eq
%--
%\bq
%\begin{split}
B_{\rm{C}}^{(1)}(x,y)
&=&
\frac{4}{\epsilon}\, \frac{x^2}{y} L_e(x)
+ 2\,
\frac{x^2}{y}\,
\Bigl[\,
-\, 8\, \zeta_2
-\, 2\, L_e(x)\,
-\,  L_e^2(y)\,
+\, 2\, L_e(x)\, L_e(y)\,
\, \Bigr] \nonumber \\
&-& 4\, ( x \,+\, y)
L_e(x)
.
\end{eqnarray}
%--
For the computation of the non-fermionic corrections these functions are needed
up to first order in $\ep$, since they are combined with the real
emission. However, this higher-order expansion is not relevant here.
\section{The Net Fermionic NNLO Differential Cross Section}\label{tot}
In this Section we use the results of Section~\ref{main}
and derive an explicit expression for the NNLO differential
cross section of Eq.~\eqref{complete}.

Note that the full set of two-loop fermionic virtual corrections to 
Bhabha scattering represents an infrared-divergent quantity.
In order to obtain a finite quantity, we take into account
the real emission of soft photons\footnote{The energy $\omega$ carried by a soft  photon in the final state
is small with respect to the center-of-mass energy $E$ introduced in Eq.~\eqref{Mandelstam}.}
from the external legs of one-loop \emph{fermionic} diagrams (class {\rm 1a}, Figure~\ref{1loop}).
The exact result is available in the literature, see e.g.
Eq.~(25) and Appendix A in~\cite{Bonciani:2004qt}.
Here we show the high-energy approximation relevant for our computation.
We consider events involving a single soft photon
carrying energy $\omega$ in the final state,
%--
\begin{equation}
e^{_-} \, (p_1)\, +\, e^{_+} \, (p_2)\,
\to\,
e^{_-} \, (p_3)\, +\, e^{_+} \, (p_4)\, + \gamma(k),
\end{equation}
%--
and compute one-loop purely-fermionic corrections.
Obviously, these real corrections factorize
and their structure is completely equivalent to the tree-level ones.
In complete analogy with Eq.~\eqref{expansion} we write
%--
\bq\label{expansionREAL}
\frac{d \sigma_{\gamma}}{d \Omega}\,=\,
\left(\frac{\alpha}{\pi}\right)
\frac{d \sigma_{\gamma}^{\rm{\ssL\ssO}}}{d \Omega}\,+\,
\left(\frac{\alpha}{\pi}\right)^2
\frac{d \sigma_{\gamma}^{\rm{\ssN\ssL\ssO}}}{d \Omega}
+ {\cal O}(\alpha^5),
\eq
%--
where
%--
\begin{eqnarray}
\label{bornREAL}
\frac{d \sigma_\gamma^{\rm{\ssL\ssO}}}{d \Omega}&=&
\frac{\alpha^2}{s}\,
\Bigl[\,
\frac{1}{2\, s^2}\,
v_1(s,t;\ep)\,
+\,
\frac{1}{2\, t^2}\,
v_1(t,s;\ep)\,
+\,  \frac{1}{s\, t}\, 
v_2(s,t;\ep)
\Bigr]\, F(\omega,s,t,m_e^2),\\
\label{NLOREAL}
\frac{d \sigma_\gamma^{\rm{\ssN\ssL\ssO}}}{d \Omega}&=&
\frac{\alpha^2}{s}
\Bigl\{
\frac{1}{s^2}\,
v_1(s,t;\ep)\,
 \sum_{f}\, Q_f^2\, \text{Re}\, \Bigl[\,   \Pi^{(1)}_f(s)\, \Bigr] \nonumber \\
&&\quad\, + \frac{1}{t^2}\, 
v_1(t,s;\ep)\,
\sum_{f}\, Q_f^2\ \text{Re}\, \Bigl[ \,   \Pi^{(1)}_f(t)\, \Bigr] \nonumber \\
&&\quad\, + \frac{1}{s\, t}\, 
v_2(s,t;\ep)\,
\,  \sum_{f}\, Q_f^2\, \text{Re}\, \Bigl[\,
 \Pi^{(1)}_f(s)\,+\, \Pi^{(1)}_f(t)\, 
\Bigr]\,\,\,
\Bigr\}\,  F(\omega,s,t,m_e^2).
\end{eqnarray}
$\Pi^{(1)}_f(x)$ can be read in Eq.~\eqref{Pi10} and, at variance with Eqs.~\eqref{born}-\eqref{NLO},
the kinematical factors introduced in Eq.~\eqref{kinCOEFF}
need to be expanded up to ${\cal O}(\ep)$,
since the real-emission
factor shows an infrared divergency,
\begin{eqnarray}
F(\omega,s,t,m_e^2)&=&
-\frac{2}{\ep}
\Bigl[
\ln\left( \frac{s}{m_e^2}\right)
+\ln\left( -\frac{t}{s}\right)
- \ln\left( 1+\frac{t}{s}\right)
-1
\Bigr]\nonumber\\
&+&\ln^2\left( \frac{s}{m_e^2}\right)+
2 \, \ln\left( \frac{s}{m_e^2}\right)
\Bigl[
2\, \ln\left(\frac{2\omega}{\sqrt{s}}\right)
+\ln\left( -\frac{t}{s}\right)
- \ln\left( 1+\frac{t}{s}\right)
\Bigr]\nonumber\\
&+ & 4 \, \ln\left(\frac{2\omega}{\sqrt{s}}\right)
\Bigl[
\ln\left( -\frac{t}{s}\right)
- \ln\left( 1+\frac{t}{s}\right) -1
\Bigr]\nonumber\\
&-& 4 \,\zeta_2
+\ln^2\left( -\frac{t}{s}\right)
- \ln^2 \left( 1+\frac{t}{s}\right)\nonumber\\
&-&2 \text{Li}_2\left( -\frac{t}{s}\right)
+2 \text{Li}_2\left( 1+\frac{t}{s}\right)
.
\end{eqnarray}

Summing the virtual contributions of Eq.~\eqref{complete} to
the real-photon emission of Eq.~\eqref{NLOREAL}  we write
the NNLO fermionic corrections to Bhabha scattering
through the sum of electron-loop contributions ($d\sigma^{\rm \ssN\ssN\ssL\ssO,e}$)
and components arising from heavier fermion loops,
\begin{eqnarray}\label{fermions}
\frac{d \sigma^{\rm{\ssN\ssN\ssL\ssO}}} {d\Omega}
+
\frac{d \sigma_\gamma^{\rm{\ssN\ssL\ssO}}} {d\Omega}
= \frac{d \sigma^{\rm{\ssN\ssN\ssL\ssO},e}} {d\Omega}
+
\sum_{f\neq e} Q_f^2 \frac{d \sigma^{\rm{\ssN\ssN\ssL\ssO},f^2}} {d\Omega}
+
\sum_{f\neq e} Q_f^4 \frac{d \sigma^{\rm{\ssN\ssN\ssL\ssO},f^4}} {d\Omega}
+
\sum_{f_1,f_2\neq e} Q_{f_1}^2 Q_{f_2}^2 \frac{d \sigma^{\rm{\ssN\ssN\ssL\ssO},2f}} {d\Omega}.
\end{eqnarray}
The double summation over the fermion species arises from the loop-by-loop
terms of Eqs.~\eqref{otherLoopByLoop} and~\eqref{loopBYloop}.
Here we do not include the case $f_1=f_2=e$, which is incorporated in
$d\sigma^{\rm \ssN\ssN\ssL\ssO,e}$.
Note also the term proportional
to $Q_f^4$, coming from Eq.~\eqref{Q4}.
The result for electron loops can be found in Eq.~(46) of~\cite{Bonciani:2004qt}.
For heavier fermion loops we introduce $x=-t \slash s$ and get:
\\
%
%------------------------------------------------------------------------------
% EQ 4.7
%------------------------------------------------------------------------------
\begin{eqnarray}
\frac{d \sigma^{\rm{\ssN\ssN\ssL\ssO},f^4}} {d\Omega}&=& \frac{\alpha^2}{2s} \Bigl\{
\frac{\left( 1-x+x^2\right)^2}{x^2}
\Bigl[ \ln\left(\frac{s}{m_e^2}\right) + \ln(R_f) + 4 \zeta_3 - \frac{5}{6}\Bigr]
+ \ln(x)
\Bigl(
\frac{1}{x^2}
-\frac{3}{2 x}
+\frac{3}{2}
- \frac{x}{2}
\Bigr) \Bigr\}, 
\\ 
%------------------------------------------------------------------------------
% EQ 4.8
%------------------------------------------------------------------------------
\frac{d \sigma^{\rm{\ssN\ssN\ssL\ssO},2f}} {d\Omega}&=&\frac{\alpha^2}{s} \Bigl\{
\frac{\left( 1-x+x^2\right)^2}{3 x^2}
\Bigl[ \ln^2\left( \frac{s}{m_e^2} \right)
+
\ln(R_{f_1})\ln(R_{f_2})+
\ln\left( \frac{s}{m_e^2} \right)
\Bigl(
\ln(R_{f_1})+\ln(R_{f_2})-\frac{10}{3}
\Bigr)\nonumber\\
&-& \frac{5}{3} 
\Bigl(
\ln(R_{f_1})+\ln(R_{f_2})-\frac{5}{3}
\Bigr)
\Bigr]
+ \frac{1}{3} \ln^2(x)
\Bigl(
\frac{1}{x^2}
- \frac{4}{3x}
+ \frac{7}{6}
- \frac{x}{3}
\Bigr)
+ \frac{\zeta_2}{3}
\Bigl(
\frac{2}{x} - 5 + 4 x - 2 x^2
\Bigr)
\nonumber\\
&+& \ln(x)
\Bigl[
\ln(R_{f_1})+\ln(R_{f_2})-\frac{10}{3}
+ 2 \ln\left( \frac{s}{m_e^2} \right)
\Bigr]
\Bigl(
\frac{1}{3x^2}-\frac{1}{2x}+\frac{1}{2}-\frac{x}{6}
\Bigr) \Bigr\}
,
\\
%------------------------------------------------------------------------------
% EQ 4.9
%------------------------------------------------------------------------------
\frac{d \sigma^{\rm{\ssN\ssN\ssL\ssO},f^2}} {d\Omega}&=& 
\frac{\alpha^2}{s}\, \Bigl\{
\sigma_1^{\rm{\ssN\ssN\ssL\ssO},f^2}  + \sigma_2^{\rm{\ssN\ssN\ssL\ssO},f^2} \,
\ln\left( \frac{2 \omega}{\sqrt{s}} \right)
\Bigr\},
\\ 
%-------------------------------------------------------------
% EQ 4.10
%------------------------------------------------------------------------------
\sigma_1^{\rm{\ssN\ssN\ssL\ssO},f^2}  
&=&
\frac{\left( 1-x+x^2\right)^2}{3 x^2}
\Bigl\{
-\frac{1}{3}
\Bigl[
\ln^3\left( \frac{s}{m_e^2}\right)
+
\ln^3\left(R_f\right)
\Bigr]
+\ln^2\left( \frac{s}{m_e^2}\right)
\Bigl[
\frac{55}{6}
- \ln\left(R_f\right)
+ \ln \left(1-x \right)
- \ln \left(x \right)
\Bigr]\nonumber\\
&+& \ln \left( \frac{s}{m_e^2}\right)
\Bigl[
- \frac{589}{18}
+\frac{37}{3} \ln\left(R_f\right)
- \ln^2\left(R_f\right)
- 2 \ln\left(R_f\right) \Bigl( \ln\left(x \right)
-\ln\left(1-x\right) \Bigr)
- 8 \text{Li}_2\left(x\right)
\Bigr]\nonumber\\
&+& \frac{4795}{108}
-\frac{409}{18} \ln\left(R_f\right)
+ \frac{19}{6}\ln^2\left(R_f\right)
- \ln^2\left(R_f\right) \Bigl( \ln\left(x \right)
- \ln\left(1-x\right) \Bigr)
- 8 \ln\left(R_f\right)\text{Li}_2\left(x\right)\nonumber\\
&+&\frac{40}{3}\text{Li}_2\left(x\right)
\Bigr\}
+ \ln \left( \frac{s}{m_e^2} \right)
\Bigl[
\zeta_2
\Bigl(
-\frac{2}{3x^2}
+\frac{4}{3x}
+ \frac{11}{2}
- \frac{23}{3} x
+ \frac{16}{3} x^2
\Bigr)
+ \ln^2\left(x\right)
\Bigl(
- \frac{1}{3 x^2}
+ \frac{17}{12 x}\nonumber\\
&-& \frac{5}{4}
- \frac{x}{12}
+ \frac{2}{3} x^2
\Bigr)
+ \ln^2\left(1-x\right)
\Bigl(
-\frac{2}{3x^2}
+\frac{11}{6x}
-\frac{5}{2}
+ \frac{11}{6} x
- \frac{2}{3} x^2
\Bigr)
+ \ln\left(x\right) \ln\left(1-x\right)
\Bigl(
\frac{2}{3x^2}\nonumber\\
&-& \frac{4}{3x}
- \frac{1}{2}
+ \frac{5}{3} x
- \frac{4}{3} x^2
\Bigr)
+ \ln\left(x\right)
\Bigl(
\frac{55}{9 x^2}
- \frac{83}{9 x}
+ \frac{65}{6}
-\frac{85}{18}x
+ \frac{10}{9} x^2
\Bigr)
+ \frac{1}{3}\ln\left(1-x\right)
\Bigl(
-\frac{10}{3 x^2}\nonumber\\
&+&\frac{31}{6 x}
-10
+\frac{31}{6}x
-\frac{10}{3}x^2
\Bigr)
\Bigr]
+ \frac{1}{3} \ln^3\left(x\right)
\Bigl(
-\frac{1}{3x^2}
+ \frac{31}{12x}
-\frac{11}{6}
-\frac{x}{6}
+\frac{x^2}{3}
\Bigr)\nonumber\\
&+&\frac{1}{3} \ln^3\left(1-x\right)
\Bigl(
-\frac{1}{3 x^2}
+ \frac{1}{x}
- \frac{4}{3}
+ x
- \frac{x^2}{3}
\Bigr)
+ \ln^2\left(x\right) \ln\left(1-x\right)
\Bigl(
-\frac{1}{3x^2}
+ \frac{1}{3x}
- \frac{4}{3}
+ x\nonumber\\
&-& \frac{x^2}{3}
\Bigr)
+ \frac{1}{3}\ln\left(x\right) \ln^2\left(1-x\right)
\Bigl(
-\frac{1}{x^2}
+ \frac{2}{x}
- \frac{7}{4}
+ \frac{x}{2}
\Bigr)
+ \ln^2\left(x\right)
\Bigl[
\frac{55}{18x^2}
- \frac{46}{9x}
+ \frac{14}{3}
- \frac{4}{9} x
- \frac{10}{9}x^2\nonumber\\
&+& \ln\left(R_f\right)\Bigl(
-\frac{1}{3 x^2}
+ \frac{17}{12 x}
- \frac{5}{4}
- \frac{x}{12}
+ \frac{2}{3} x^2
\Bigr)
\Bigr]
+ \ln^2\left(1-x\right)
\Bigl[
\frac{10}{9 x^2}
-\frac{29}{9 x}
+ \frac{9}{2}
-\frac{29}{9}x
+ \frac{10}{9} x^2\nonumber\\
&+& \ln\left(R_f\right)\Bigl(
-\frac{2}{3x^2}
+ \frac{11}{6x}
- \frac{5}{2}
+ \frac{11}{6}x
- \frac{2}{3}x^2
\Bigr)
\Bigr]
+ \ln\left(x\right)\ln\left(1-x\right)
\Bigl[
- \frac{10}{9x^2}
+ \frac{37}{18x}
+ \frac{1}{2}
- \frac{25}{9}x\nonumber\\
&+& \frac{20}{9}x^2
+ \ln\left(R_f \right)\Bigl(
\frac{2}{3x^2}
- \frac{4}{3x}
- \frac{1}{2}
+ \frac{5}{3}x
- \frac{4}{3} x^2
\Bigr)
\Bigr]
+ \ln\left(x\right)
\Bigl[
-\frac{589}{54 x^2}
+ \frac{1753}{108 x}
- \frac{701}{36}
+ \frac{925}{108}x\nonumber\\
&-& \frac{56}{27}x^2
+\text{Li}_2\left(x\right)
\Bigl(
-\frac{4}{x^2}
+ \frac{19}{3x}
- 7
+ 3 x
- \frac{2}{3}x^2
\Bigr)
+ \ln\left(R_f\right)
\Bigl(
\frac{37}{9x^2}
- \frac{56}{9x}
+ \frac{47}{6}
- \frac{67}{18}x
+ \frac{10}{9}x^2
\Bigr)\nonumber\\
&+& \zeta_2
\Bigl(
-\frac{2}{3x^2}
+ \frac{4}{x}
-\frac{1}{6}
- \frac{10}{3} x
+2 x^2
\Bigr)
\Bigr]
+ \ln\left(1-x\right)
\Bigl[
\frac{56}{27x^2}
- \frac{161}{54x}
+\frac{56}{9}
- \frac{161}{54}x
+ \frac{56}{27}x^2\nonumber\\
&+&\ln\left(R_f\right)
\Bigl(
-\frac{10}{9x^2}
+ \frac{31}{18x}
-\frac{10}{3}
+ \frac{31}{18}x
-\frac{10}{9}x^2
\Bigr)
+\zeta_2 \Bigl(
-\frac{2}{x^2}
+ \frac{20}{3x}
-\frac{32}{3}
+ \frac{20}{3}x
-2x^2
\Bigr)
\Bigr]\nonumber\\
&+&\text{Li}_3\left(x\right)
\Bigl(
\frac{4}{3x^2}
-\frac{7}{3x}
+ 3
-\frac{5}{3}x
+\frac{2}{3}x^2
\Bigr)
+ \frac{2}{3}S_{1,2}\left(x\right)
\Bigl(
-\frac{1}{x^2}
+\frac{1}{x}
-x
+ x^2
\Bigr)\nonumber\\
&+&\zeta_2
\Bigl[
\frac{19}{9x^2}
-\frac{13}{18x}
-\frac{43}{3}
+\frac{311}{18}x
-\frac{98}{9}x^2
+\ln\left(R_f\right)
\Bigl(
-\frac{2}{3x^2}
+\frac{4}{3x}
+\frac{11}{2}
-\frac{23}{3}x
+\frac{16}{3}x^2
\Bigr)
\Bigr]\nonumber\\
&+& \zeta_3
\Bigl(
-\frac{4}{3x^2}
+\frac{3}{x}
- 5
+\frac{11}{3}x
-2x^2
\Bigr)
,
\\ 
%----------------------------------------------------------
% EQ 4.11
%------------------------------------------------------------------------------
\sigma_2^{\rm{\ssN\ssN\ssL\ssO},f^2}
&=&
\frac{8}{3} \frac{\left( 1-x+x^2\right)^2}{x^2}
\Bigl\{
\ln^2\left( \frac{s}{m_e^2}\right)
+ \ln \left( \frac{s}{m_e^2}\right)
\Bigl[
-\frac{8}{3}
+ \ln \left(R_f\right)
- \ln \left(1-x\right)
\Bigr]
+ \ln \left(x \right) \ln \left( R_f \right)\nonumber\\
&+& \Bigl[
\frac{5}{3}-\ln\left( R_f \right)
\Bigr]
\Bigl[
1 + \ln \left( 1-x \right)
\Bigr]
\Bigr\}\nonumber\\
&+& 4 \Bigl[ \ln\left( \frac{s}{m_e^2}\right) 
\ln(x)
\Bigl(
\frac{4}{3x^2}
- \frac{7}{3x}
+ 3
- \frac{5}{3}x
+ \frac{2}{3}x^2
\Bigr) 
+  \ln^2(x)
\Bigl(
\frac{2}{3 x^2}
- \frac{1}{x}
+ 1
- \frac{1}{3}x
\Bigr)\nonumber\\
&-&  \ln(x)\ln(1-x)
\Bigl(
 \frac{2}{3 x^2}
- \frac{1}{x}
+ 1
-\frac{1}{3}x
\Bigr)
- \frac{1}{3} \ln(x)
\Bigl(
 \frac{16}{3 x^2}
- \frac{29}{3 x}
+ 13
-\frac{23}{3}x
+ \frac{10}{3}x^2\Bigr) \Bigr]
.
\end{eqnarray}
%----------------------------------
%% definition of S_{n,p}
%----------------------------------
In order to have compact results we used
\begin{equation}
S_{n,p}\left(y\right)
=
\frac{(-1)^{n+p-1}}{(n-1)!p!}\int_{0}^1 \, dx\, \frac{\ln^{n-1}(x)\ln^p(1-xy)}{x}.
\end{equation}

\begin{figure}[t]
%\begin{center}
%\begin{minipage}[c]{.48\textwidth}
\includegraphics[scale=0.7]{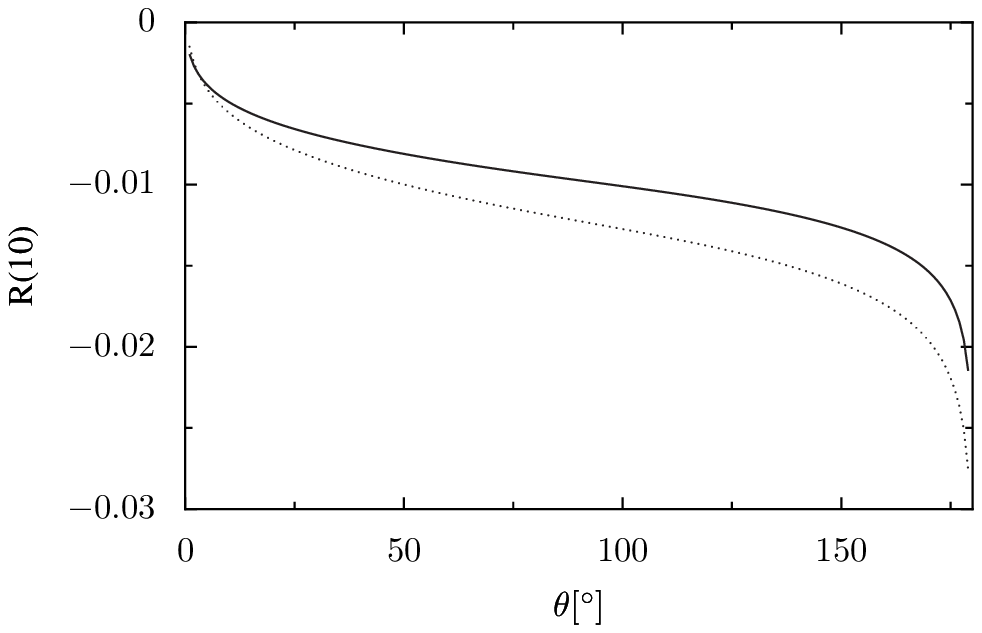}
%\end{minipage}
%\ 
%\begin{minipage}[c]{.48\textwidth}
\includegraphics[scale=0.7]{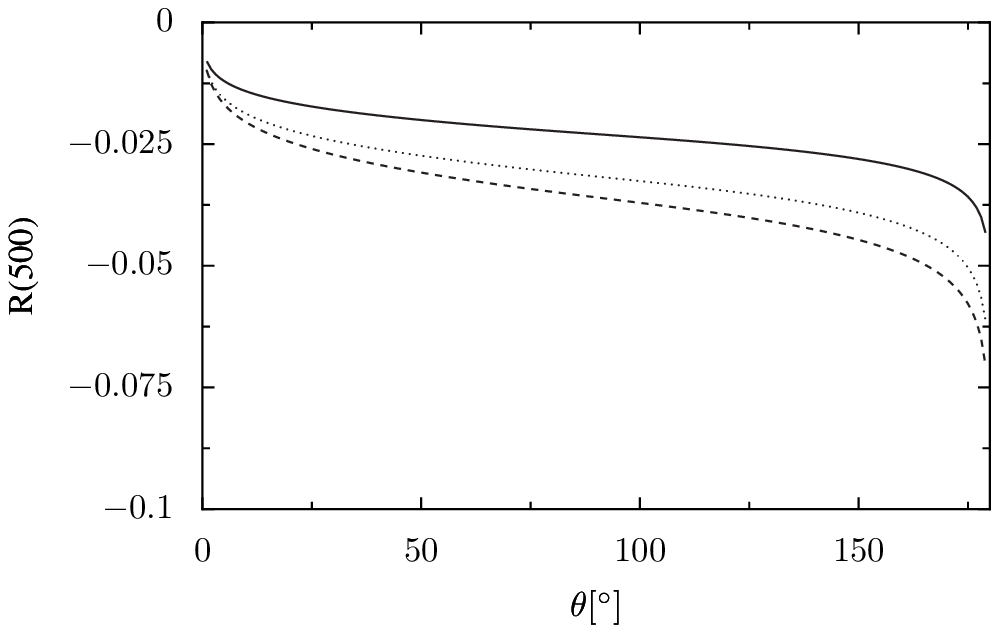}
%--
%\end{minipage}
%\end{center}
\caption{Ratio of the fermionic NNLO corrections to the differential cross section 
respect to the tree-level result for $\sqrt{s}=10$ GeV and $\sqrt{s}=500$ GeV.
A solid line represents the electron-loop contributions,
a dotted one the sum of electron- and muon-loop ones,
and a dashed one includes also $\tau$ leptons.}
\label{2plots}
\end{figure}

In Table~\ref{3Degs} (Table~\ref{90Degs}) we show numerical values for the NNLO
corrections to the differential cross section for a scattering angle $\theta=3^\circ$ ($\theta=90^\circ$).
In both tables we set  $\omega=E \slash 10$.
Finally, in Figure~\ref{2plots} we plot the ratio of the two-loop fermionic
corrections to the tree-level cross section,
\begin{equation}
R(\sqrt{s})\, =\,
\left( \frac{\alpha}{\pi} \right)^2\,
\frac{d \sigma^{\rm{\ssN\ssN\ssL\ssO}}+ d \sigma_\gamma^{\rm{\ssN\ssL\ssO}}} {d \sigma^{\rm{\ssL\ssO}}}
\end{equation}
for $\sqrt{s}=10$ GeV and $\sqrt{s}=500$ GeV.

\begin{table}[ht]\centering
\setlength{\arraycolsep}{\tabcolsep}
\renewcommand\arraystretch{1.2}
\begin{tabular}{|l|c|l|l|}
\hline
d$\sigma$ $\slash$ d$\Omega$ [nb] $\vert$ $\sqrt{s}$\ [GeV]    & 10             & 91   & 500 \\
\hline 
\hline
LO QED \quad \quad \qquad [Eq.~\eqref{born}] &      440873          & 5323.91   & 176.349 \\
\hline 
LO \texttt{Zfitter}\, \qquad ~\cite{Bardin:1999yd,Arbuzov:2005ma} & 440875              & 5331.5  & 176.283   \\
\hline 
NNLO ($e$) \quad \, \qquad [Eq.~\eqref{fermions}] & -1397.35              & -35.8374    & -1.88151 \\
\hline 
NNLO ($e\, +\, \mu$)\, \, \qquad \ `` &  -1394.74              & -43.1888 & -2.41643  \\
\hline 
NNLO ($e\, +\, \mu \, +\tau$) \, \, `` &                &   & -2.55179 \\
\hline 
NNLO photonic  \, \quad ~\cite{Glover:2001ev,Penin:2005eh} & 9564.09               & 251.661 & 12.7943 \\
\hline 
\end{tabular}
\caption[]{Numerical values for the NNLO corrections to the differential cross section respect to the solid angle.
Results are expressed in nanobarns for a scattering angle $\theta=3^\circ$.
Empty entries are related to cases where the high-energy approximation cannot be
applied.}
\label{3Degs}
\end{table}
%--
\begin{table}[ht]\centering
\setlength{\arraycolsep}{\tabcolsep}
\renewcommand\arraystretch{1.2}
\begin{tabular}{|l|c|l|l|}
\hline 
d$\sigma$ $\slash$ d$\Omega$ [nb] $\vert$ $\sqrt{s}$\ [GeV]    & 10             & 91   & 500 \\
\hline
\hline
LO QED \quad \quad \qquad [Eq.~\eqref{born}] &      0.466409          & 0.00563228   & 0.000186564 \\
\hline 
LO \texttt{Zfitter}\, \qquad ~\cite{Bardin:1999yd,Arbuzov:2005ma} &  0.468499    & 0.127292  &  0.0000854731    \\
\hline 
NNLO ($e$) \quad \, \qquad [Eq.~\eqref{fermions}] &  -0.00453987  &  -0.0000919387   & -4.28105 $\cdot$ $10^{-6}$ \\
\hline 
NNLO ($e\, +\, \mu$)\, \, \qquad \ `` &    -0.00570942            & -0.000122796  &-5.90469 $\cdot$ $10^{-6}$  \\
\hline 
NNLO ($e\, +\, \mu \, +\tau$) \, \, `` &  -0.00586082 &  -0.000135449  & -6.7059 $\cdot$ $10^{-6}$ \\
\hline 
NNLO ($e\, +\, \mu \, +\tau +\, t$) \, \, `` &   &   & -6.6927 $\cdot$ $10^{-6}$ \\
\hline 
NNLO photonic  \, \quad ~\cite{Glover:2001ev,Penin:2005eh} &  0.0358755    &0.000655126 &  0.0000284063\\
\hline 
\end{tabular}
\caption[]{Numerical values for the NNLO corrections to the differential cross section respect to the solid angle.
Results are expressed in nanobarns for a scattering angle $\theta=90^\circ$.
Empty entries are related to cases where the high-energy approximation cannot be
applied.}
\label{90Degs}
\end{table}
%--

\begin{figure}[t]
%\begin{center}
%\begin{minipage}[c]{.48\textwidth}
\includegraphics[scale=0.7]{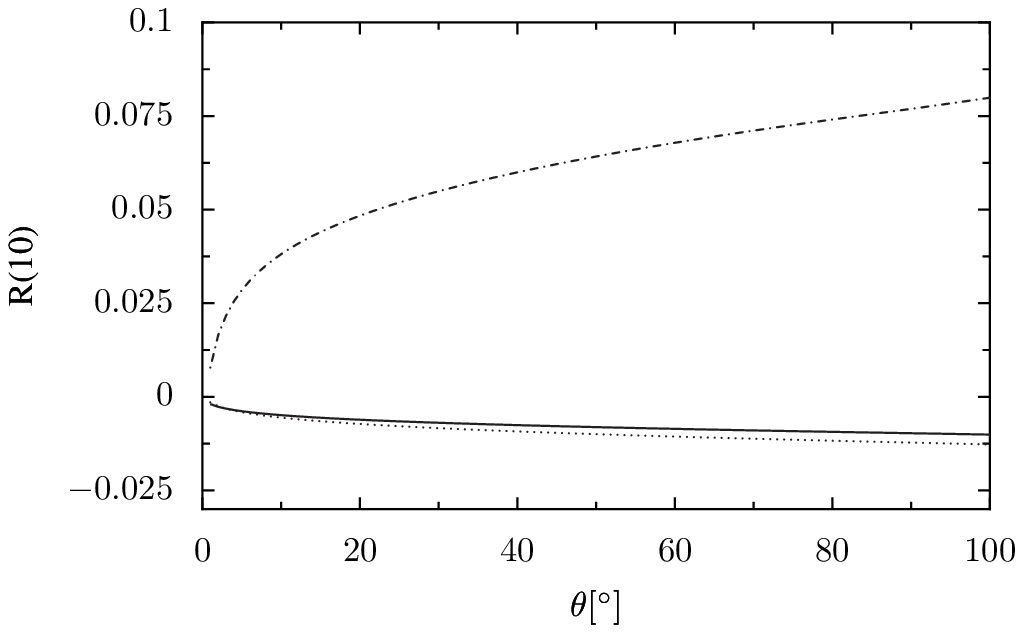}
%\end{minipage}
%\ 
%\begin{minipage}[c]{.48\textwidth}
\includegraphics[scale=0.7]{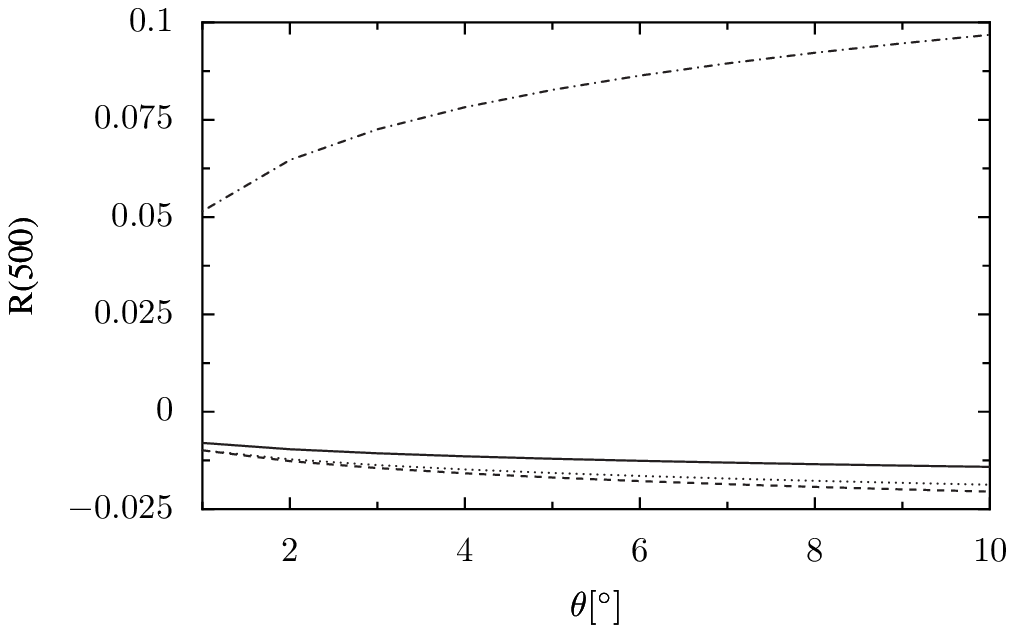}
%--
%\end{minipage}
%\end{center}
\caption{Same as Figure~\ref{2plots}, including the photonic
contributions of~\cite{Arbuzov:1998du,Glover:2001ev,Penin:2005eh} (dash-dotted lines).}
\label{2plotsScan}
\end{figure}

It is clear from the Tables, that although there is no decoupling of the
heavier fermions (as indeed there shouldn't, since the typical scale of the
process is large compared to all the masses), the electron loop contributions
dominate in the fermionic part and the latter is still substantially smaller
than the pure photonic corrections.
% tr 04042007
%-------------------------
\section{Summary}
\label{final}
%-------------------------
In this article, we completed the computation of the virtual two-loop QED fermionic
corrections to Bhabha scattering.
Based on the kinematics of the targeted phenomenological applications, we
considered the limit $m_e^2 \ll m_f^2 \ll s,t,u$.

The fermionic double box contributions with two different mass scales have
been derived for the first time here.  Their numerical importance is
comparable to the two-loop self-energies and vertices. We note, however, a
qualitative difference. Due to the structure of the collinear singularities of
the graphs, the contributions of the heavier fermions are not suppressed.

A numerical estimation of differential cross sections shows that the net
fermionic two-loop effects may be neglected for applications at LEP 1 and LEP
2, but have to be taken into account for precision calculations when a level
of $10^{-4}$ has to be reached, as is anticipated for the Giga-Z option of the
ILC project.
 
Completing the NNLO program for Bhabha scattering requires still several ingredients.
First, let us mention the contributions
from the five light quark flavors.  Here, an approach based on dispersion
relations {\it \`a la} \cite{Kniehl:1988id} should be suitable.  On the other hand, the
heavy top quark might be considered decoupling in a large part of the
interesting kinematical regions. 
Furthermore,  an implementation of the loop-by-loop corrections
with pentagon diagrams has to be done. 

Finally, light fermionic pair emission diagrams need to be considered. As
known from the form-factor case, they are responsible for the cancellation
of the leading part of the logarithmic sensitivity on the masses.
\\
Exact and approximated results are made publicly available at~\cite{webPage:2006xx}.
The combination of our result with the photonic two-loop corrections
of~\cite{Penin:2005eh} and with electron loop corrections of ~\cite{Bonciani:2004gi,Bonciani:web} proves well-suited for phenomenogical purposes,
e.g. a precise luminosity determination at a future International
Linear Collider.
%-------------------------
\section*{Acknowledgements}
%-------------------------
We would like to thank A.~Arbuzov, R.~Bonciani, A.~Ferroglia and
A.~Penin  for useful communications, and S.~Moch and A.~Mitov for interesting
dicussions.

Work supported in part 
%------
by 
Sonderforschungsbereich/Transregio TRR 9 of DFG 
`Computerge\-st{\"u}tz\-te Theo\-re\-ti\-sche Teil\-chen\-phy\-sik',  
%-----
by
the Sofja Kovalevskaja Award of the Alexander von Humboldt Foundation
  sponsored by the German Federal Ministry of Education and Research,
%-----
by
the ToK Program ``ALGOTOOLS'' (MTKD-CD-2004-014319)
%-----
by 
the Polish State Committee for Scientific Research (KBN),
research projects in 2004--2005, 
%-----
and by 
the European Community's Marie-Curie Research Training Networks  MRTN-CT-2006-035505 `HEPTOOLS'
%`Tools and Precision Calculations for Physics Discoveries at Colliders'
%
and MRTN-CT-2006-035482 `FLAVIAnet'.
%-------------------------
\section*{Note added}
%-------------------------
We would like to thank T.~Becher and K.~Melnikov for drawing our
attention to a problem with a first version of our result, which lead
us to discover an incorrectly expanded integral. After correction, Eq.
4.9 agrees with the result published in the meantime in~\cite{Becher:2007cu}.

%%%%%%%%%%%%%%%%%%%%%%%%%%%%%%%%%%%%%%%%%%%%%%%%%%%%%%%%%%%%%%%%%%%%%%%%%%%%%%%%%%%%%%%
\appendix
\section{\label{appendix}Mass-Expanded Master Integrals}
%--
The list of Master Integrals required by our computation
can already be found in Table V of~\cite{Czakon:2004wm}.
The eight most difficult masters, those involving two different
mass scales, have been derived in~\cite{Actis:2006dj}.
%(Eqs.(7)-(8) for \texttt{SE3l2M1m(d)}, Eqs.(14)-(15) for \texttt{V4l2M1m(d)},
%Eqs.(16)-(17) for \texttt{V4l2M2m(d)} and Eqs.(21)-(22) for \texttt{B5l2M2m(d)}).
Because they are a substantial part of the present study we reproduce them here: 
\begin{eqnarray}
%--
\texttt{SE3l2M1m[on shell]} &=&  M^2\ m^{-4 \epsilon} 
\Bigl\{ 
R\Bigl[
\frac{1}{2\epsilon^2}
+\frac{5}{4\epsilon}
-\frac{3}{8}
+\frac{\zeta_2}{2}
+\frac{3}{2}L(R)
-\frac{1}{2} L^2(R)
\Bigr]\nonumber\\
&+&R^2 \Bigl[
\frac{11}{18}-\frac{1}{3}L(R)
\Bigr]
+\epsilon
\Bigl[
R\Bigl(
\frac{45}{16} + \frac{5}{4}\zeta_2
-\frac{\zeta_3}{3} -\frac{7}{4} L(R)
+ L^2(R)\nonumber\\ 
&-&\frac{1}{2}L^3(R)
\Bigr)
+ R^2 \Bigl(
-\frac{3}{4} + \frac{8}{9} L(R) -\frac{1}{2}L^2(R)
\Bigr)
\Bigr]
%\frac{1}{\epsilon^2}
%+ \frac{1}{\epsilon}
%\Bigl[
%3 + 2 \ L \left(R\right)
%\Bigr]
%+
%7 + \zeta_2  + 6 \ L \left(R\right)
%+ 2 \ L^2 \left(R\right)\nonumber\\
%&+& \epsilon
%\Bigl[
%15 + 3\zeta_2
%  - \frac{2 \zeta_3}{3} + \left(14 + 2\zeta_2\right) 
% L \left(R\right)
%+ 6 L^2 \left(R\right) +\frac{4}{3} L^3 \left(R\right)
%\Bigr]
 \Bigr\}, \\
&&\nonumber\\
\texttt{SE3l2M1md[on shell]} &=&
m^{-4 \epsilon}
\Bigl\{ 
\frac{1}{2 \epsilon^2}
+ 
\frac{1}{2\epsilon} \Bigl[ 1+ 2 L \left(R\right)\Bigr]
+
\frac{1}{2}\left(1 + \zeta_2\right) + L \left(R\right) + L^2 \left(R\right)\nonumber\\
&+& \epsilon
\Bigl[
\frac{1}{6} \left(3 + 3\zeta_2 - 2\zeta_3\right)
+ \left(1 + \zeta_2\right)
L \left(R\right) 
+L^2 \left(R\right) + \frac{2}{3} L^3 \left(R\right)
\Bigr]\nonumber\\
&+&R
\Bigl[
-\frac{3}{4}
+ \frac{1}{2} L(R)
+ \epsilon \left(
\frac{7}{8}- L(R) + \frac{3}{4}L^2(R)
\right)
\Bigr]\nonumber\\
&+& R^2 \Bigl[ 
-\frac{5}{36}
+ \frac{1}{6}L(R)
+ \ep  \Bigl(
-\frac{5}{72}
+ \frac{1}{18}L(R)
+ \frac{1}{4}L^2(R)
\Bigr) \Bigr]
\Bigr\}, \\
&&\nonumber\\
\texttt{V4l2M1m[x]} &=&
m^{-4 \epsilon} 
\Bigl\{ 
\frac{1}{2\epsilon^2}
+ \frac{5}{2\epsilon}
+ \frac{1}{2}\Bigl[ 19 - 3\zeta_2 - L_m^2(x) \Bigr]\nonumber\\
&+&\frac{M^2}{x}\Bigl[
-2 + 4 \zeta_2 - 4 \zeta_3 -2 L_m(x)
+ 2 L_M(x)
-4 \zeta_2 L_M(x)\nonumber\\
&+& 2 L_m(x) L_M(x)
- L_M^2(x)
- L_m(x) L_M(x)
+ L_M^3(x)
\Bigr]
\Bigr\},\\
&&\nonumber\\
\texttt{V4l2M1md[x]} &=& 
 \frac{m^{-4 \epsilon}}{m^2}
\Bigl\{  
\frac{1}{2\epsilon^2}
+ \frac{1}{\epsilon}
\Bigl[
1+\frac{1}{2} L_m(x)
\Bigr]
+
2-\zeta_2 +L_m(x)+\frac{1}{4}\, L_m^2(x)\nonumber\\
&+&\frac{M^2}{x}\Bigl[
\frac{1}{\epsilon}
-\frac{1}{\epsilon}L_M(x)
- 1 + 3 \zeta_2 + L_m(x) +L_M(x)\nonumber\\
&-& L_m(x)L_M(x)-\frac{1}{2}L_M^2(x)
\Bigr]
\Bigr\},\\
&&\nonumber\\
\texttt{V4l2M2m[x]} &=&
m^{-4 \epsilon} 
\Bigl\{
\frac{1}{2\epsilon^2}
+
\frac{1}{\epsilon}
\Bigl[
\frac{5}{2} +L_m(x)
\Bigr]
+
\frac{1}{2}(19+\zeta_2) +  5\ L_m(x)
+ L_m^2(x)
\Bigr\},\\
&&\nonumber\\%----------------------------------------------
\texttt{V4l2M2md[x]} &=&
 \frac{m^{-4 \epsilon}}{6 x} 
\Bigl[ 12\zeta_3 - 6\zeta_2 L_M(x) - L_M^3(x) \Bigr],\\
&&\nonumber \\\label{b5l2m2m}%----------------------------------------------
\texttt{B5l2M2m[x,y]} &=&
\frac{m^{-4 \epsilon}}{x} 
\Bigl\{
\frac{1}{\epsilon^2}L_m(x)
+\frac{1}{\epsilon}
\Bigl(
-\zeta_2
+2 L_m(x)
+ \frac{1}{2}L_m^2(x)
+ L_m(x)L_m(y)
\Bigr)\nonumber\\
&-&2 \zeta_2
- 2 \zeta_3
+ 4 L_m(x)
+ L_m^2(x)
+ \frac{1}{3} L_m^3(x)
- 4 \zeta_2 L_m(y)\nonumber\\
&+& 2 L_m(x)L_m(y)
+ L_m(x)L_m^2(y)
-\frac{1}{6}L_m^3(y)\nonumber\\
&-& \Bigl( 3 \zeta_2 
+ \frac{1}{2} L_m^2(x)
- L_m(x) L_m(y) 
+ \frac{1}{2} L_m^2(y) \Bigr) \ln\left( 1+ \frac{y}{x}\right)\nonumber\\
&-& \Bigl( L_m(x) -L_m(y) \Bigr) \text{Li}_2 \left(-\frac{y}{x}\right)
+ \text{Li}_3 \left(-\frac{y}{x}\right)
\Bigr\},\\
&&\nonumber\\%----------------------------------------------
\texttt{B5l2M2md[x,y]} &=&
\frac{m^{-4 \epsilon}}{xy} 
\Bigl\{ 
\frac{1}{\epsilon}
\Bigl[
-L_m(x)L_m(y)+L_m(x)L(R)
\Bigr]
-2 \zeta_3
+ \zeta_2 L_m(x)
+ 4 \zeta_2 L_m(y)\nonumber\\
&-& 2 L_m(x)L^2_m(y)
+ \frac{1}{6}L^3_m(y)
- 2 \zeta_2 L(R)
+ 2 L_m(x) L_m(y)L(R)- \frac{1}{6}L^3(R)\nonumber\\
&+& \Bigl( 3 \zeta_2  + \frac{1}{2} L_m^2(x) -  L_m(x) L_m(y)
+ \frac{1}{2} L_m^2(y) \Bigr) \ln\left( 1+ \frac{y}{x}\right)\nonumber\\
&+& \Bigl( L_m(x) -L_m(y) \Bigr) \text{Li}_2 \left(-\frac{y}{x}\right)
- \text{Li}_3 \left(-\frac{y}{x}\right)
 \Bigr\}. 
%--
\end{eqnarray}
We list also the other \emph{expanded} masters, including the correct
normalizations. Note that, compared to the conventions
employed in~\cite{Czakon:2004wm} and in Eq.~\eqref{selfOneOld}, all integrals
are rescaled by a factor $m^{L(D-2l)}$, where
$L$ is the number of loops, $D=4-2 \ep$ and $l$ is
the number of internal lines.
Expansions are performed up to the order required
by our computation. For example, we include ${\cal O}(m^2)$ terms in \texttt{SE2l2m[x]} (see Eq.~\eqref{selfOne})
since the reduction procedure generates coefficients containing $1 \slash m^2$.
The same consideration applies to ${\cal O}(\ep)$ terms,
which are included as long as the reduction brings inverse powers
of $\ep$ in the coefficient functions.
Since in the following no ambiguities
arise, in we drop the subscript $f$ and we set $L(x)=\ln (-m^2/x)$,

% taken from: BHABHA_QED/Masters/MATH_COMMON/MastBhabhaEpsExp.m

\begin{eqnarray}
%MTm[m]=((m^2*(48*(1 + ep + ep^2 + ep^3 + ep^4) + 24*ep^2*(1 + ep + ep^2)*z2 
%    + 27*ep^4*z4 + 16*ep^3*z3*(-1 - ep + 2*ep*Log[m]) + 
%   8*ep*Log[m]*(-6*(1 + ep)*(2 + ep^2*(2 + z2)) + 
%   ep*Log[m]*(12 + 6*ep*(2 + ep*(2 + z2)) +
%	      4*ep*Log[m]*(-2*(1 + ep) + ep*Log[m])))))/(48*ep) +ep^4*TadEps);
\texttt{T1l1m}\,=\, (m^2)^{1-\ep}\, \Bigl[
\frac{1}{\ep}\, +\,
1\, +\,
\ep\, \Bigl( 
1+\frac{\zeta_2}{2}
\Bigr)
+
\ep^2
\Bigl(
1+\frac{\zeta_2}{2}-\frac{\zeta_3}{3}
\Bigr)
\Bigr],
\end{eqnarray}

\begin{eqnarray}\label{selfOne}
%SE2l2mExp = (1/ep + 
%     2 + Log[ms] + ms*(-2 + 2*Log[ms])
%    + ep* (ms*(-2*(1 + z2) + Log[ms]^2) + (8 - z2 + 4*Log[ms] +
%                 Log[ms]^2)/2)
%    +ep^2* (8 - z2 - (7*z3)/3 - ms*(2 + z2 + 4*z3) -
%    ((-8 + z2 + 2*ms*z2)*Log[ms])/2 + Log[ms]^2 + ((1 + 2*ms)*Log[ms]^3)/6)
%     + ep^3*SE2l2mEps);
\texttt{SE2l2m[x]} &=& m^{-2 \ep}\, \Bigl\{
\frac{1}{\ep}\,
+\, 2 \, +\, L(x) \, +\, 2\,  \frac{m^2}{x}\, \Bigl[\, 1\,-\,L(x)\,\Bigr]\, +\, \ep\,
\Bigl[
\, 4\, -\, \frac{\zeta_2}{2}\, +\, 2\,L(x) \nonumber \\ 
&+&\, \frac{1}{2}\,  L^2(x)\, +\, \frac{m^2}{x}\,
\Bigl(\,
2\, +\,2\, \zeta_2\, -\, L^2(x)
\, \Bigr)\,
\Bigr]
\, +\, \ep^2\,
\Bigl[\,
8\,-\,\zeta_2\, -\, \frac{7}{3}\,\zeta_3 \nonumber \\ 
&+& 4\, L(x)
- \frac{1}{2}\,\zeta_2\,L(x)\, +\, L^2(x)\, +\, \frac{1}{6}\,L^3(x)\, +\, \frac{m^2}{x}\,
\Bigl(\,
2\, +\, \zeta_2 \nonumber\\ &+& 4 \,\zeta_3 \,+\, \zeta_2\, L(x)\, -\, \frac{1}{3}\, L^3(x)
\Bigr)
\Bigr]
\Bigr\},
\end{eqnarray}

\begin{eqnarray}
% SE2l0mExp[s] = ((-6*(1 + 2*ep)*(-2 + ep^2*(-8 + z2)) - 28*ep^3*z3 + 
%  ep*Log[-s]*(-12 + 6*ep*(-4 + ep*(-8 + z2)) + 
%	      2*ep*Log[-s]*(3 + 6*ep - ep*Log[-s])))/(12*ep)+ep^3*SE2l0mEps);
\texttt{SE2l0m[x]} &=& m^{-2\ep} \Bigl\{
\frac{1}{\ep}\,+\,2\,+\,L(x)\,
+\,\ep\,
\Bigl[\,
4\,-\,\frac{\zeta_2}{2}\,+\,2\,L(x)\,+\,\frac{1}{2}\,L^2(x)\,
\Bigr] \nonumber \\
&+&\ep^2
\Bigl[
8-\zeta_2\,-\,\frac{7}{3}\,\zeta_3\,+\,4\,L(x)\,-\,\frac{1}{2}\,\zeta_2\,L(x)\,
+\,L^2(x) + \frac{1}{6}\,L^3(x)\,
\Bigr]\Bigr\},
\end{eqnarray}

\begin{eqnarray}
%V3l1mExp = (-(ms*(24*z2 + 30*ep*z3 - 6*ep*z2*Log[ms] + 3*Log[ms]^2 + 
%		  ep*Log[ms]^3))/6 + ep^2*V3l1mEps);
\texttt{V3l1m[x]} = \frac{m^{-2\ep}}{x}\,
\Bigl\{
4\, \zeta_2 + \frac{1}{2}\, L^2(x)\,-\,
\ep
\Bigl[
-5\,\zeta_3 + \zeta_2\, L(x)-\frac{1}{6}\, L^3(x)
\Bigr]
\Bigr\},
\end{eqnarray}

\begin{eqnarray}
\texttt{SE3l1m[on shell]}&=& (m^2)^{1-2\ep}\Bigl[
\frac{1}{2\, \ep^2}\, +\, \frac{5}{4\, \ep}\,+\,\frac{11}{8}\,
+\,\frac{5}{2}\,\zeta_2\,
+\,\ep\,\Bigl(\,
-\,\frac{55}{16}\,+\,\frac{25}{4}\,\zeta_2\,+\,\frac{11}{3}\,\zeta_3\, \Bigr)\nonumber \\
&+& \ep^2\, \Bigl(\,
-\,\frac{949}{32}\,+\,\frac{55}{8}\,\zeta_2\,+\,\frac{55}{6}\,\zeta_3\,+\,\frac{303}{8}\,\zeta_4\,
\Bigr)\Bigr],
\end{eqnarray}

\begin{eqnarray}
%SE3l2MExp = (ep^2*SE3l2mEps + 
% (M^2*(48*Ms + ep*(12 + 144*Ms + ep*(78 + 8*Ms*(30 + 6*z2) + 
%        ep*(345 - 12*z2 + 24*Ms*(6 + 6*z2)))) + 
%    24*ep^2*(ep + 4*Ms + 12*ep*Ms)*Log[M^2]^2 - 
%    64*ep^3*Ms*Log[M^2]^3 - 4*ep*Log[M^2]*
%     (24*Ms + ep*(6 + 72*Ms + ep*(39 + 4*Ms*(30 + 6*z2))) + 
%      12*ep^2*Log[Ms]*(1 - 2*Ms*Log[Ms])) + 
%    4*ep^2*(64*ep*Ms*z3 + Log[Ms]*(6 + ep*(39 + 4*Ms*(-12 + 6*z2)) + 
%        6*Log[Ms]*(ep - 2*Ms - 6*ep*Ms - 2*ep*Ms*Log[Ms])))))/
%	     (48*ep^2*Ms));
\texttt{SE3l2m[x]}&=&(m^2)^{1-2\ep}\Bigl\{
\frac{1}{\ep^2} + \frac{1}{\ep} \Bigl( 3-\frac{x}{4 m^2} \Bigr)
+5+\zeta_2-L^2(x)-\frac{x}{m^2} \Bigl[ \frac{13}{8} + \frac{1}{2}L(x)\Bigr] \nonumber \\
&+&\ep \,\Bigl[\,
3\, +\, 3\, \zeta_2\, +\, \frac{16}{3}\,\zeta_3\, -\, 4\, L(x)\,
+\,2\,\zeta_2\,L(x)\, -\, 3\, L^2(x)\,-\,L^3(x)\nonumber\\
&-&\frac{x}{m^2}\,\Bigl(\,
\frac{115}{16}\, -\, \frac{\zeta_2}{4}\,+\,\frac{13}{4}\,L(x)\,+\,\frac{1}{2}\,L^2(x)\,
\Bigr)\,
\Bigr] \Bigr\},
\end{eqnarray}

\begin{eqnarray}
%SE3l2MdExp = ((
%ep^2*SE3l2mdEps + 
% (120 + ep*(120 - 120*ep*(1 + ep*(11 + 49*ep)) + 11*ep^3*Pi^4 + 
%     120*ep*(1 + 3*ep + 9*ep^2)*z2) - 
%   10*ep*(16*ep^2*(1 + ep)*Log[M^2]^3 - 8*ep^3*Log[M^2]^4 - 
%     4*ep*Log[M^2]^2*(6 + ep*(6 + ep*(-6 + 6*z2)) - 
%       6*ep^2*Log[Ms]*(2 + Log[Ms])) + 4*Log[M^2]*
%      (6 + ep*(6 + ep*(-6 + 6*z2 + 3*ep*(-22 + 6*z2))) + 32*ep^3*z3 - 
%       2*ep^2*Log[Ms]*(6 - ep*(-30 + 6*z2) + 3*Log[Ms]*
%          (1 + 4*ep + ep*Log[Ms]))) + 
%     ep*(-64*ep*z3*(1 + 4*ep + 3*ep*Log[Ms]) + 
%       Log[Ms]*(24 - 4*ep*(-30 + 6*z2 + 3*ep*(-38 + 6*z2)) + 
%         Log[Ms]*(12 + 48*ep - 4*ep^2*(-42 + 6*z2) + 
%		  ep*Log[Ms]*(12 + 40*ep + 7*ep*Log[Ms]))))))/(240*ep^2))
%	      +
%(-2*Ms - 2*ep^2*Ms - (13*ep^2*Ms*Pi^4)/180 + 2*ep^2*Ms*z2 - 
% 6*ep*Ms*z3 - 6*ep^2*Ms*z3 - 2*ep*Ms*(2 + z2 + ep*z2 + 8*ep*z3)*
%  Log[Ms] + Ms*(1 + ep - ep^2*(3 + 2*z2))*Log[Ms]^2 + 
% ep*(1 + ep)*Ms*Log[Ms]^3 + (7*ep^2*Ms*Log[Ms]^4)/12 + 
% 2*ep^2*Ms*Log[M^2]^2*(-2 + Log[Ms]^2) + 
% 2*ep*Ms*Log[M^2]*(2 + 6*ep*z3 + 2*ep*(2 + z2)*Log[Ms] - 
%		   (1 + ep)*Log[Ms]^2 - ep*Log[Ms]^3) )
%	      )
%;
\texttt{SE3l2md[x]}&=&m^{-4\ep}\Bigl\{
\frac{1}{2\ep^2}+\frac{1}{2\ep} - \frac{1}{2} + \frac{\zeta_2}{2}
- L(x)-\frac{1}{2}L^2(x)
-\frac{m^2}{x} \Bigl[ -2 + L^2(x)\Bigr]\nonumber\\
&+& \ep\,
\Bigl[
-\frac{11}{2}+\frac{3}{2}\zeta_2 + \frac{8}{3}\zeta_3 - 5 L(x)
+\zeta_2 L(x)
-2 L^2(x)-\frac{1}{2} L^3(x)\nonumber \\
&-&
\frac{m^2}{x}\, \Bigl(
-6 \, \zeta_3 \,-\,4 \,L(x)\,
-2\, \zeta_2\,L(x)\, +\, L^2(x)\,+\,L^3(x)
\Bigr)
\Bigr] \Bigr\}.
\end{eqnarray}
%
%\begin{eqnarray}
%B4l2mExpms = ((2*ms^2*Log[ms])/(ep*x) + 
% (ms^2*(12*Log[ms]^2 - 12*Log[ms]*Log[x]))/(6*x) + 
% (ep*ms^2*(24*z3 - 54*z2*Log[ms] + 4*Log[ms]^3 + 
%    36*z2*Log[x] - 6*Log[ms]^2*Log[x] + 2*Log[x]^3 - 
%    36*z2*Log[1 + x] + 12*Log[-x]*Log[x]*Log[1 + x] - 
%    6*Log[x]^2*Log[1 + x] + 12*Log[x]*
%	   PolyLog[2, 1 + x] + 12*PolyLog[3, -x]))/(6*x)+ ep^2*B4l2mEpsS); 
%\end{eqnarray}
%\begin{eqnarray}
%test equation
%\end{eqnarray}
Finally, the mass expanded one-loop box Master Integral \texttt{B4l2m[x,y]} can be collected from Eqs.(4.70)-(4.75) 
of~\cite{Fleischer:2006ht}:
\begin{eqnarray}
\texttt{B4l2m[x,y]} &=&
\frac{m^{-2\epsilon}}{xy}
\Bigl\{
\frac{2}{\epsilon}
\Bigl[
L(y)- \ln\left( \frac{x}{y}\right)
\Bigr]
+
2 L^2(y) - 2L(y)\ln\left( \frac{x}{y}\right)
+\epsilon 
\Bigl[
4 \zeta_3
- 9 \zeta_2 L(y)\nonumber\\
&+& \frac{2}{3} L^3(y)
+ 5 \zeta_2 \ln\left( \frac{x}{y}\right)
- L^2(y) \ln \left( \frac{x}{y} \right)
+ \frac{1}{3} \ln^3\left( \frac{x}{y}\right)
- 6 \zeta_2 \ln\left( 1 + \frac{x}{y} \right)\nonumber\\
&+& 2 \ln\left( - \frac{x}{y} \right)
\ln\left( \frac{x}{y}\right)
\ln\left( 1 + \frac{x}{y}\right)
- \ln^2 \left( \frac{x}{y} \right)
\ln\left( 1 + \frac{x}{y}\right)\nonumber\\
&+& 2 \ln\left( \frac{x}{y} \right)
\text{Li}_2\left( 1 + \frac{x}{y} \right)
+ 2 \text{Li}_3\left( - \frac{x}{y} \right)
\Bigr]
\Bigr\}.
\end{eqnarray}

%
%\bibliographystyle{utphys_spires}
%\bibliography{2loops}

\providecommand{\href}[2]{#2}

\end{document}